\documentclass[a4paper,11pt]{article}
\pdfoutput=1 

\usepackage{jheppub} 

\usepackage{mathrsfs}
\usepackage{bbm}
\usepackage{bm}
\usepackage{array,longtable}
\usepackage{multirow}
\usepackage{slashed}
\usepackage[utf8]{inputenc}

\allowdisplaybreaks

\title{\bf \boldmath $CP$ asymmetry in the angular distributions of $\tau\to K_S\pi\nu_\tau$ decays~--~II: general effective field theory analysis}

\author[a]{Feng-Zhi Chen,}
\author[b,1]{Xin-Qiang Li,\note{Corresponding author.}}
\author[b]{Shi-Can Peng,}
\author[b]{Ya-Dong Yang,}
\author[a,1]{Hong-Hao Zhang}

\affiliation[a]{School of Physics, Sun Yat-Sen University, Guangzhou 510275, China,}
\affiliation[b]{Institute of Particle Physics and Key Laboratory of Quark and Lepton Physics~(MOE), Central China Normal University, Wuhan, Hubei 430079, China}

\emailAdd{chenfzh25@mail.sysu.edu.cn}
\emailAdd{xqli@mail.ccnu.edu.cn}
\emailAdd{shicanpeng@mails.ccnu.edu.cn}
\emailAdd{yangyd@mail.ccnu.edu.cn}
\emailAdd{zhh98@mail.sysu.edu.cn}

\abstract{In this work, we proceed to study the $CP$ asymmetry in the angular distributions of $\tau\to K_S\pi\nu_\tau$ decays within a general effective field theory framework including four-fermion operators up to dimension-six. It is found that, besides the commonly considered scalar-vector interference, the tensor-scalar interference can also produce a non-zero $CP$ asymmetry in the angular distributions, in the presence of complex couplings. Using the dispersive representations of the $K\pi$ form factors as inputs, and taking into account the detector efficiencies of the Belle measurement, we firstly update our previous SM predictions for the $CP$ asymmetries in the same four $K\pi$ invariant-mass bins as set by the Belle collaboration. Bounds on the effective couplings of the non-standard scalar and tensor interactions are then obtained under the combined constraints from the $CP$ asymmetries measured in the four bins and the branching ratio of $\tau^-\to K_S\pi^-\nu_\tau$ decay, with the numerical results given respectively by $\mathrm{Im}[\hat{\epsilon}_S]=-0.008\pm0.027$ and $\mathrm{Im}[\hat{\epsilon}_T]=0.03\pm0.12$, at the renormalization scale $\mu_\tau=2~\mathrm{GeV}$ in the $\mathrm{\overline{MS}}$ scheme. Using the best-fit values, we also find that the distributions of the $CP$ asymmetries can deviate significantly from the SM expectation in almost the whole $K\pi$ invariant-mass region. Nevertheless, the current bounds on $\mathrm{Im}[\hat{\epsilon}_S]$ and $\mathrm{Im}[\hat{\epsilon}_T]$ are still plagued by large experimental uncertainties, but will be improved with more precise measurements from the Belle II experiment as well as the proposed Tera-Z and STCF facilities. Assuming further that the non-standard scalar and tensor interactions originate from a weakly-coupled heavy new physics well above the electroweak scale, the $SU(2)_L$ invariance of the resulting SMEFT Lagrangian would indicate that very strong limits on $\mathrm{Im}[\hat{\epsilon}_S]$ and $\mathrm{Im}[\hat{\epsilon}_T]$ could also be obtained from the neutron electric dipole moment and the $D^0-\bar{D}^0$ mixing. With the bounds from these processes taken into account, it is then found that, unless there exist extraordinary cancellations between the new physics contributions, neither the scalar nor the tensor interaction can produce any significant effects on the $CP$ asymmetries (relative to the SM predictions) in the processes considered, especially under the ``single coefficient dominance'' assumption.}

\begin{document} 
\maketitle
\flushbottom

\section{Introduction}
\label{sec:intro}

The violation of charge-conjugation and parity-reversal ($CP$) symmetry (CPV) is a necessary condition for explaining the observed matter-antimatter asymmetry in the Universe~\cite{Sakharov:1967dj}. To date, CPV in weak interactions has been experimentally established in the quark sector through $K$-, $B$- and $D$-meson decays~\cite{Christenson:1964fg,Burkhardt:1988yh,AlaviHarati:1999xp,Fanti:1999nm,Aubert:2001nu,Abe:2001xe,Aubert:2004qm,Chao:2004mn,Aaij:2012kz,Aaij:2013iua,Aaij:2019kcg}, and all the phenomena could be generally accommodated within the Standard Model (SM) by the single irreducible complex phase present in the Cabibbo-Kobayashi-Maskawa (CKM) quark-mixing matrix~\cite{Cabibbo:1963yz,Kobayashi:1973fv}. However, the SM description of CPV fails to accommodate the observed baryon asymmetry of the Universe, entailing therefore additional sources of CPV beyond the SM. An attractive alternative to the problem is via the so-called leptogenesis mechanism~\cite{Fukugita:1986hr}, in which CPV is driven by leptodynamics. Indeed, the evidence for neutrino oscillations implies that $CP$ could also be violated in the lepton sector~\cite{Xing:2019vks,Branco:2011zb}, as indicated by the recent T2K result~\cite{Abe:2019vii}. Among the six lepton species, the $\tau$ lepton is especially compelling as it is massive enough to decay into either light leptons or hadrons, implying that a host of decay channels are available to be studied. Interestingly, the hadronic $\tau$ decays, besides serving as a clean laboratory to study the low-energy aspect of strong interaction~\cite{Pich:2013lsa,Davier:2005xq}, may also be a good place to study CPV both within the SM and beyond~\cite{Tsai:1996ps,Kuhn:1996dv,Bigi:2012km,Bigi:2012kz,Kiers:2012fy,Roig:2019rwf}.

Within the SM, the hadronic $\tau$ decays proceed via the exchanges of $W^{\pm}$ bosons and, as the CKM matrix elements involved are real and the strong phases must be the same in the two $CP$-conjugated processes, there exists no direct CPV in these decays. Nevertheless, when the well-established $CP$ asymmetry in $K^0-\bar{K}^0$ mixing is taken into account, a non-zero indirect CPV could still arise in the processes involving a $K_S$ or a $K_L$ meson in the final state~\cite{Bigi:2005ts}. Therefore, any significant excess of $CP$ asymmetry beyond the SM expectation can be served as a clear hint of new physics (NP). Assuming that the hadronic $\tau$ decays receive an additional contribution from some NP, which carries different weak and strong phases from that of the SM term, one can then construct $CP$-violating observables in terms of the interference between the SM and NP amplitudes. Being of linear dependence on the potential NP amplitude, these observables show a higher sensitivity to NP than do other SM-forbidden ones, such as the $\tau\to \mu\gamma$ decay rate and the electric dipole moment (EDM) of leptons, which are usually quadratic in the NP amplitude~\cite{Bigi:2012km,Bigi:2012kz}.

In this work, we will focus on the CPV in $\tau\to K_S\pi\nu_\tau$ decays, which has been searched for by several experiments. After the initial null results from CLEO~\cite{Anderson:1998ke,Bonvicini:2001xz} and Belle~\cite{Bischofberger:2011pw}, a non-zero $CP$ asymmetry was reported for the first time by the BaBar collaboration~\cite{BABAR:2011aa}, by measuring the decay-rate difference between $\tau^+$ and $\tau^-$ decays. However, such a measurement is in conflict with the SM prediction~\cite{Bigi:2005ts,Calderon:2007rg,Grossman:2011zk,Chen:2019vbr} at the level of $2.8\sigma$, which has motivated many NP explanations by including the extra contribution from non-standard tensor interactions~\cite{Cirigliano:2017tqn,Rendon:2019awg,Chen:2019vbr,Dighe:2019odu,Devi:2013gya,Dhargyal:2016kwp,Dhargyal:2016jgo,vonDetten:2021euh}. Unfortunately, the suppression of the relative phase between the $K\pi$ vector and tensor form factors as well as the combined constraints from other relevant observables have already excluded such a possibility~\cite{Cirigliano:2017tqn,Rendon:2019awg,Chen:2019vbr}. Although it has been shown in Ref.~\cite{Dighe:2019odu} that, using a gauge-invariant dimension-eight tensor operator, one can account for the $CP$ anomaly while evading the most stringent bound from the neutron EDM and keeping at the same time the extraction of $|V_{us}|$ from exclusive $\tau$ decays unaffected, such a scenario would induce a host of other issues, such as the breakdown of the power counting of the underlying effective field theory (EFT) as well as the constraints from the spectrum of $\tau^-\to K_S\pi^-\nu_\tau$ decay itself and the semi-leptonic kaon decays~\cite{Roig:2019rwf}. Thus, further precise measurements of the decay-rate asymmetry from the Belle II experiment~\cite{Kou:2018nap} as well as the proposed Tera-Z~\cite{Pich:2020qna} and Super Tau Charm Facility (STCF)~\cite{Sang:2020ksa} is essential to make any conclusive statement about the observed $CP$ anomaly.

In fact, from a theoretical point of view, the decay-rate asymmetry itself is not an ideal observable aimed for $CP$ studies due to the following observations. As only the vector-tensor interference contributes to the decay-rate asymmetry~\cite{Devi:2013gya}, the $K\pi$ vector and tensor form factors, which encode the hadronization information of the corresponding quark currents, play a key role in determining the amount of direct $CP$ asymmetry in the decay. While the $K\pi$ vector form factor has been well studied~\cite{Boito:2008fq,Boito:2010me,Jamin:2006tk,Jamin:2008qg,Escribano:2014joa,Bernard:2009zm,Bernard:2011ae,Bernard:2013jxa,Moussallam:2007qc,Kimura:2009pm,Kimura:2014wsa,Beldjoudi:1994hi,Epifanov:2007rf}, this is obviously not the case for the tensor form factor due to the lack of sufficient data on the tensor interactions, and we have to construct it merely from theory. To this end, a convenient approach is to use the dispersion relation, which warrants simultaneously the properties of unitarity and analyticity, with the phase input for the dispersive representation obtained using the chiral theory with resonances (R$\chi$T)~\cite{Ecker:1988te,Ecker:1989yg}, together with the theoretical constraints at both low and high energies~\cite{Cirigliano:2017tqn,Rendon:2019awg,Chen:2019vbr}. However, as the same spin-1 resonances contributing to the $K\pi$ vector form factor will equivalently contribute to the tensor form factor~\cite{Ecker:1988te,Ecker:1989yg}, the phases of the two form factors must be the same as long as the $K\pi$ states dominate the unitarity relation and, according to the Watson's final-state interaction theorem~\cite{Watson:1954uc}, should coincide with the phase shift of the isospin-$\frac{1}{2}$ P-wave $K\pi$ scattering amplitude in the elastic region (\textit{i.e.}, from the $K\pi$ threshold up to $\sim 1.2~\mathrm{GeV}$)~\cite{Cirigliano:2017tqn}. This implies the exact absence of direct CPV in the elastic region, and leaves only the inelastic region, where the Watson's theorem is invalid, to be possible to have a non-zero contribution to the direct CPV. However, only the asymptotic behaviour rather than the explicit information of the $K\pi$ tensor form factor is currently known in the inelastic region, rendering therefore the amount of $CP$ asymmetry induced by a tensor operator not only strongly suppressed but also very uncertain~\cite{Cirigliano:2017tqn,Rendon:2019awg,Chen:2019vbr}. On the other hand, due to the lack of vector-scalar interference in the decay-rate asymmetry, any potential direct CPV induced by the non-standard scalar interactions cannot be probed through such an observable. Therefore, a more suitable observable exempt from the aforementioned defects is urgently called for. 

For this purpose, an interesting observable is the $CP$ asymmetry in the angular distributions of $\tau\to K_S\pi\nu_\tau$ decays, which can be measured for unpolarized single $\tau$'s even if their rest frame cannot be reconstructed~\cite{Kuhn:1996dv}. Following the same notation as adopted in Ref.~\cite{Kou:2018nap},\footnote{It should be mentioned that the variable $\theta$, which is defined as the angle between the direction opposite to that of the $e^+e^-$ center-of-mass (CM) system and the direction of the hadronic system in the $\tau$ rest frame, used by Belle~\cite{Bischofberger:2011pw} is not adopted by Belle II~\cite{Kou:2018nap}, since such an angle is relevant only when the $\tau$ polarization is known. Here our notation is consistent with that of Belle II, and we will use the variable $\theta$ only when using the Belle data in the numerical analysis.} we can write the $CP$-violating observable as
\begin{equation}\label{eq:ACP_i}
A_i^{CP}=\frac{\int_{s_{1,i}}^{s_{2,i}}\int_{-1}^{1} \cos\alpha \left[\frac{d^2 \Gamma(\tau^-\to K_S\pi^-\nu_\tau)}{ds\,d\cos\alpha} -\frac{d^2 \Gamma(\tau^+\to K_S\pi^+\bar{\nu}_\tau)}{ds\,d\cos\alpha}\right] ds\,d\cos\alpha}{\frac{1}{2}\int_{s_{1,i}}^{s_{2,i}}\int_{-1}^{1}\left[\frac{d^2 \Gamma(\tau^-\to K_S\pi^-\nu_\tau)}{ds\,d\cos\alpha}+\frac{d^2 \Gamma(\tau^+\to K_S\pi^+\bar{\nu}_\tau)}{ds\,d\cos\alpha}\right] ds\,d\cos\alpha}\,,
\end{equation}
which is defined as the difference between the differential $\tau^-$ and $\tau^+$ decay widths weighted by $\cos\alpha$, with $\alpha$ being the angle between the directions of $K$ and $\tau$ in the $K\pi$ rest frame,\footnote{Here $\cos\alpha=\cos\beta\cos\psi+\sin\beta\sin\psi\cos\phi$, where, relative to the laboratory direction (chosen as $+z$) in the $K\pi$ rest frame, the direction of $K$ is characterized by the polar angle $\beta$ and the azimuthal angle $\phi$, while the direction of $\tau$ by the polar angle $\psi$ and its relative azimuthal angle is fixed such that $\tau$ is located in the $y$-$z$ plane. Note that the direction of $\tau$ cannot be measured due to the missing neutrino, but the polar angle $\psi$ can be determined in terms of the energy of the hadronic system in the laboratory frame~\cite{Kuhn:1996dv,Kou:2018nap}.} and can be evaluated in different bins of the $K\pi$ invariant mass squared $s$, with the $i$-th bin given by the interval $[s_{1,i}, s_{2,i}]$~\cite{Bischofberger:2011pw}. Three of us have pointed out for the first time that, as a $K_S$ meson is involved in the final state, the well-established CPV in $K^0-\bar{K}^0$ mixing can induce a non-zero $CP$ asymmetry in the angular distributions even within the SM~\cite{Chen:2020uxi}. However, our predictions are still below the current Belle detection sensitivity of $\mathcal{O}(10^{-3})$ and lie within the margins of the Belle results measured in four different bins of the $K\pi$ invariant mass, except for a $1.7\sigma$ deviation for the lowest bin~\cite{Chen:2020uxi}. This, on the one hand, needs to be verified by future precise measurements with higher sensitivity by the Belle II experiment~\cite{Kou:2018nap} and, on the other hand, leaves room for possible NP contributions.

Direct CPV in the angular distributions of $\tau\to K_S\pi\nu_\tau$ decays can be induced by the interference between the S-wave from exotic scalar-exchange and the P-wave from SM $W$-exchange diagrams, provided that the couplings of exotic scalars to fermions are complex, and has been studied for both polarized and unpolarized beams~\cite{Tsai:1996ps,Kuhn:1996dv}. Although such a scenario has already been considered in a number of specific NP models~\cite{Tsai:1996ps,Kuhn:1996dv,Kimura:2009pm,Kimura:2014wsa}, a general model-independent analysis is still missing. In addition, as the amplitude of $W$-exchange diagram can be decomposed into both a P-wave and an S-wave part, there exists actually another source of direct CPV in the angular distributions, due to the interference between a non-standard tensor interaction with complex couplings and the S-wave term from the $W$-exchange diagram. This feasibility provides a new avenue to probe the non-standard tensor interaction and has the advantage over the decay-rate asymmetry, since one can now avoid the uncertainty brought by the $K\pi$ tensor form factor by setting the $K\pi$ invariant-mass intervals within the elastic region, where the explicit information of the form factor is quite clear due to the Watson's theorem~\cite{Watson:1954uc}. It is even reliable to extend the analysis to the inelastic region, since the relative phase between the tensor and scalar form factors is large enough to make the impact of the form-factor uncertainties less important.

Without loss of generality, we will perform here a model-independent analysis of the $CP$ asymmetry in the angular distributions of $\tau\to K_S\pi\nu_\tau$ decays, within a low-energy EFT framework including the most general four-fermion interactions among the SM fields up to dimension-six. As a key ingredient for predicting the amount of $CP$ asymmetry in the processes considered, both the moduli and the phases of $K\pi$ vector, scalar and tensor form factors are needed. To this end, we will use as inputs the dispersive representations~\cite{Boito:2008fq,Boito:2010me,Jamin:2006tk,Jamin:2008qg,Jamin:2000wn,Jamin:2001zq,Jamin:2006tj,Cirigliano:2017tqn,Rendon:2019awg,Chen:2019vbr} rather than the Breit-Wigner parametrizations~\cite{Epifanov:2007rf,Paramesvaran:2009ec,Finkemeier:1995sr,Finkemeier:1996dh} of these form factors, because the former can warrant the properties of unitarity and analyticity and contain a full knowledge of QCD in both the perturbative and non-perturbative regimes, while the latter do not vanish at threshold and even violate the Watson's theorem before the higher resonances come into play~\cite{Cirigliano:2017tqn,Rendon:2019awg,Chen:2019vbr}. It should be mentioned that our previous SM predictions~\cite{Chen:2020uxi} are obtained by following the same notation as specified in Ref.~\cite{Kou:2018nap} and, in order to infer the allowed NP contributions from the available Belle data~\cite{Bischofberger:2011pw}, we must firstly update the SM predictions by adapting to the Belle environment and taking into account the detector efficiencies of the Belle measurement~\cite{Bischofberger:2011pw}. Bounds on the effective couplings of the non-standard interactions can then be obtained under the combined constraints from the $CP$ asymmetries measured in four $K\pi$ invariant-mass bins~\cite{Bischofberger:2011pw} and the branching ratio of $\tau^-\to K_S\pi^-\nu_\tau$ decay~\cite{Epifanov:2007rf,Zyla:2020zbs}, with the numerical results given respectively by $\mathrm{Im}[\hat{\epsilon}_S]=-0.008\pm0.027$ and $\mathrm{Im}[\hat{\epsilon}_T]=0.03\pm0.12$, at the renormalization scale $\mu_\tau=2~\mathrm{GeV}$ in the modified minimal subtraction ($\mathrm{\overline{MS}}$) scheme. Using the obtained best-fit values, we will also present the distributions of the $CP$ asymmetries, which are found to deviate significantly from the SM prediction in almost the whole $K\pi$ invariant-mass region. While being still plagued by large experimental uncertainties, the current constraints will be improved with more precise measurements from the Belle II experiment~\cite{Kou:2018nap}, as well as the future Tera-Z~\cite{Pich:2020qna} and STCF~\cite{Sang:2020ksa} facilities. Assuming further that the non-standard scalar and tensor interactions originate from a weakly-coupled heavy NP well above the electroweak scale, the $SU(2)_L$ invariance of the resulting Standard Model Effective Field Theory (SMEFT) Lagrangian~\cite{Buchmuller:1985jz,Grzadkowski:2010es,Brivio:2017vri} would indicate that very strong limits on $\mathrm{Im}[\hat{\epsilon}_S]$ and $\mathrm{Im}[\hat{\epsilon}_T]$ can also be obtained from the neutron EDM and the $D^0-\bar{D}^0$ mixing~\cite{Cirigliano:2017tqn}. With the bounds from the neutron EDM and the $D^0-\bar{D}^0$ mixing taken into account, it is then found that, unless there exist extraordinary cancellations between the NP contributions, neither the scalar nor the tensor interaction can produce any significant effects on the $CP$ asymmetries (relative to the SM prediction) in the processes considered, especially under the ``single coefficient dominance'' assumption. 

The rest of this paper is organized as follows. In section~\ref{sec:CPV}, taking into account properly the $K^0-\bar{K}^0$ mixing effect, we derive the general formula for the $CP$ asymmetry in the angular distributions of $\tau\to K_S\pi\nu_\tau$ decays. In section~\ref{sec:Effective}, we present a detailed analysis of the $\tau\to K_S\pi\nu_\tau$ decays within the most general low-energy EFT framework, and obtain the explicit expression of the $CP$ asymmetry in the presence of non-standard scalar and tensor interactions. Our numerical results and discussions are then presented in section~\ref{sec:numerical}. Our conclusions are finally made in section~\ref{sec:conclusion}. For convenience, the dispersive representations of the $K\pi$ vector, scalar, and tensor form factors are given in the appendix.  

\section{\boldmath{$CP$} asymmetry in the angular distributions of \boldmath{$\tau\to K_S\pi\nu_\tau$ decays}}\label{sec:CPV}

We now recapitulate the derivation of the $CP$-violating angular observable in $\tau\to K_S\pi\nu_\tau$ decays, including contributions from both the SM and beyond; for more details, the readers are referred to Refs.~\cite{Tsai:1996ps,Kuhn:1996dv,Chen:2020uxi}. According to the well-known $\Delta S=\Delta Q$ rule, a $\tau^+$ ($\tau^-$) decays initially into the flavour eigenstate $K^0=\bar{s}d$ ($\bar{K}^0=s\bar{d}$), which then evolves into a superposition of both $K^0$ and $\bar{K}^0$ due to $K^0-\bar{K}^0$ mixing. However, the experimentally reconstructed kaons are the mass ($K_S$ and $K_L$) rather than the flavour ($K^0$ and $\bar{K}^0$) eigenstates, which are related to each other via\footnote{Here we have assumed that $CP$ is violated, while the $CPT$ invariance still holds in $K^0-\bar{K}^0$ mixing.}
\begin{align}\label{eq:mixing_cp}
&|K_{S,L}\rangle=p\,|K^0\rangle \pm q\,|\bar{K}^0\rangle\,,
\end{align}
with the normalization $|p|^2+|q|^2=1$. It is also important to realize that the $K_S$ state in $\tau\to K_S\pi\nu_\tau$ decays is not observed directly in experiment, but rather reconstructed in terms of a $\pi^+\pi^-$ final state with its invariant mass fixed around $M_K$ and by requiring the time difference between the $\tau$ and the kaon decay to be around the $K_S$ lifetime~\cite{Grossman:2011zk}. Due to the well-established CPV in $K^0-\bar{K}^0$ mixing, however, the same $\pi^+\pi^-$ pair can be obtained not only from $K_S$, but also inevitably from $K_L$, when the decay time difference is longer than the $K_L$ lifetime. Therefore, we are actually facing the cascade decays, $\tau^{\pm}\to K_{S,L}(\to\pi^+\pi^-)\pi^\pm\bar{\nu}_\tau(\nu_\tau)$, in which the initial states $\tau^{\pm}$ decay firstly into the intermediate states $K_S$ and $K_L$ that, after a time $t$, decay further into the final state $\pi^+\pi^-$. 

An intuitive and convenient way for describing the processes involving a $K_{S(L)}$ intermediate state is to use the reciprocal basis~\cite{Sachs:1963zz,Enz:1965tr,Wolfenstein:1970wb,Beuthe:1997fu,AlvarezGaume:1998yr,Silva:2000db,Silva:2004gz}. Following the steps detailed already in Ref.~\cite{Chen:2020uxi}, one can eventually obtain the time-dependent $CP$ asymmetry in the angular distributions of $\tau\to K_S\pi\nu_\tau$ decays:
\begin{align}\label{eq:ACPi}
A_i^{CP}(t_1,t_2)
=\frac{\left(\langle\cos\alpha\rangle_i^{\tau^-}+\langle\cos\alpha\rangle_i^{\tau^+}\right)A^{CP}_K(t_1,t_2)+\left(\langle\cos\alpha\rangle_i^{\tau^-}-\langle\cos\alpha\rangle_i^{\tau^+}\right)}{1+A^{CP}_K(t_1,t_2)\, A_{\tau,i}^{CP}}\,,
\end{align}
with
\begin{align}
\langle\cos\alpha\rangle_i^{\tau^-}+\langle\cos\alpha\rangle_i^{\tau^+}&=\frac{\int_{s_{1,i}}^{s_{2,i}}\int_{-1}^{1} \cos \alpha \left[\frac{d \Gamma^{\tau^-}}{d\omega}+\frac{d \Gamma^{\tau^+}}{d\omega}\right]d\omega}{\frac{1}{2}\int_{s_{1,i}}^{s_{2,i}}\int_{-1}^{1} \left[\frac{d \Gamma^{\tau^-}}{d\omega}+\frac{d \Gamma^{\tau^+}}{d\omega}\right]d\omega}\,,\label{eq:cossum}\\[0.2cm]
\langle\cos\alpha\rangle_i^{\tau^-}-\langle\cos\alpha\rangle_i^{\tau^+}&=\frac{\int_{s_{1,i}}^{s_{2,i}}\int_{-1}^{1} \cos \alpha \left[\frac{d \Gamma^{\tau^-}}{d\omega}-\frac{d \Gamma^{\tau^+}}{d\omega}\right]d\omega}{\frac{1}{2}\int_{s_{1,i}}^{s_{2,i}}\int_{-1}^{1} \left[\frac{d \Gamma^{\tau^-}}{d\omega}+\frac{d \Gamma^{\tau^+}}{d\omega}\right]d\omega}\,,\label{eq:cosdif}\\[0.2cm]
A^{CP}_K(t_1,t_2)&=\frac{\int_{t_1}^{t_2}F(t)\left[\bar\Gamma_{\pi^+\pi^-}(t)-\Gamma_{\pi^+\pi^-}(t)\right]dt}{\int_{t_1}^{t_2}F(t)\left[\bar\Gamma_{\pi^+\pi^-}(t)+\Gamma_{\pi^+\pi^-}(t)\right]dt}\,,\label{eq:AcpK}\\[0.2cm]
A_{\tau,i}^{CP}&=\frac{\int_{s_{1,i}}^{s_{2,i}}\int_{-1}^{1} \left[\frac{d \Gamma^{\tau^-}}{d\omega}-\frac{d \Gamma^{\tau^+}}{d\omega}\right]d\omega}{\int_{s_{1,i}}^{s_{2,i}}\int_{-1}^{1} \left[\frac{d \Gamma^{\tau^-}}{d\omega}+\frac{d \Gamma^{\tau^+}}{d\omega}\right]d\omega}\,,\label{eq:Acptau}
\end{align}
where $\frac{d\Gamma^{\tau^\pm}}{d\omega}=\frac{d^2\Gamma(\tau^\pm\to K^0(\bar K^0)\pi^\pm\bar\nu_\tau(\nu_\tau))}{ds\,d\cos\alpha}$, $\Gamma(\bar\Gamma)_{\pi^+\pi^-}(t)=\Gamma(K^0(\bar K^0)(t)\to\pi^+\pi^-)$, and $\langle\cos\alpha\rangle_i^{\tau^{\pm}}$ denote the differential $\tau^{\pm}$ decay widths weighted by $\cos\alpha$ and evaluated in the $i$-th bin of the $K\pi$ invariant mass squared. As $A^{CP}_K(t_1,t_2)$ (and hence $A_i^{CP}(t_1,t_2)$) is sensitive to the experimental cuts, its theoretical prediction can be made only when the kaon decay time interval $[t_1,t_2]$ over which the observable is integrated and the function $F(t)$ introduced to parametrize the experiment-dependent effects are known. While $F(t)$ should be determined as part of the experimental analysis, we do not have such a function for the moment and will quote the particularly simple prediction made in Ref.~\cite{Grossman:2011zk},\footnote{For simplicity, from now on, we will use the abbreviations $A^{CP}_K$ and $A_{(\mathrm{SM}),i}^{CP}$ to stand respectively for $A^{CP}_K(t_1\ll\Gamma_S^{-1},\Gamma_S^{-1}\ll t_2 \ll\Gamma_L^{-1})$ and $A_{(\mathrm{SM}),i}^{CP}(t_1\ll\Gamma_S^{-1},\Gamma_S^{-1}\ll t_2 \ll\Gamma_L^{-1})$, where $\Gamma_{S}$ and $\Gamma_{L}$ denote the decay widths of the short- ($K_S$) and long-lived ($K_L$) mass eigenstates.}
\begin{equation}\label{eq:ACPK}
A^{CP}_K(t_1\ll\Gamma_S^{-1},\Gamma_S^{-1}\ll t_2 \ll\Gamma_L^{-1})\approx -2\mathrm{Re}(\epsilon_K)=-(3.32\pm0.06)\times10^{-3}\,,
\end{equation}
where the kaon decay-time interval is chosen to include contributions from both the pure $K_S$ decay term and the interference term between the $K_S$ and $K_L$ decays~\cite{Grossman:2011zk}, and $\epsilon_K$ is the $CP$-violating parameter in neutral kaon decays~\cite{Zyla:2020zbs}. As pointed out already in Refs.~\cite{Chen:2019vbr,Chen:2020uxi}, several reasonable approximations and a double-step form of $F(t)$~\cite{Grossman:2011zk} have been applied to obtain Eq.~\eqref{eq:ACPK}. This might be, however, not always the case in experiment. For instance, when the efficiency function provided by BaBar is used, the SM prediction of the decay-rate asymmetry (see Eq.~\eqref{eq:intACP} for its definition) turns out to be $(3.6\pm0.1)\times10^{-3}$~\cite{BABAR:2011aa}. Nevertheless, it is still reasonable to use Eq.~\eqref{eq:ACPK} in our numerical analysis, keeping however in mind that the theoretical predictions made in this work can be further refined once the explicit form of the function $F(t)$ is provided by experiment.

Within the SM, one has $\frac{d \Gamma^{\tau^+}}{d\omega}=\frac{d \Gamma^{\tau^-}}{d\omega}$, and thus $A_{\tau,i}^{CP}=0$ and $\langle\cos\alpha\rangle_i^{\tau^-}=\langle\cos\alpha\rangle_i^{\tau^+}$. As a consequence, the $CP$ asymmetry in the angular distributions of $\tau\to K_S\pi\nu_\tau$ decays defined by Eq.~\eqref{eq:ACPi} reduces to~\cite{Chen:2020uxi}
\begin{align}\label{eq:reduceACPi}
A_{\mathrm{SM},i}^{CP}=2\,\langle\cos\alpha\rangle_i^{\tau^-}A^{CP}_K\,,
\end{align}
which implies that, once $A^{CP}_K$ is fixed by Eq.~\eqref{eq:ACPK}, $A_{\mathrm{SM},i}^{CP}$ will be determined exclusively by $\langle\cos\alpha\rangle_i^{\tau^-}$. For a detailed computation of $\langle\cos\alpha\rangle_i^{\tau^-}$ as well as the SM predictions of $A_{\mathrm{SM},i}^{CP}$ in several bins of the $K\pi$ invariant mass, the readers are referred to Ref.~\cite{Chen:2020uxi}. It should be emphasized again that such a non-zero $CP$ asymmetry in the angular distributions is induced by the well-established CPV in $K^0-\bar{K}^0$ mixing~\cite{Chen:2020uxi}. 

In the presence of NP contributions to $\tau^\pm\to K^0(\bar K^0)\pi^\pm\bar\nu_\tau(\nu_\tau)$ decays, however, it is generally expected that $\frac{d \Gamma^{\tau^+}}{d\omega}\neq\frac{d \Gamma^{\tau^-}}{d\omega}$, and thus both Eqs.~\eqref{eq:cosdif} and \eqref{eq:Acptau} would deviate from zero. Moreover, Eq.~\eqref{eq:cossum} will be modified by the additional NP contributions. As a consequence, one may obtain different values of $A_i^{CP}$ with respect to the SM expectations. Since the second term is expected to be much smaller than the first one in the denominator of Eq.~\eqref{eq:ACPi}, the $CP$ asymmetry $A_i^{CP}$ can be further approximated by
\begin{align}\label{eq:ACPNP}
A_i^{CP}\simeq \left(\langle\cos\alpha\rangle_i^{\tau^-}+\langle\cos\alpha\rangle_i^{\tau^+}\right)A^{CP}_K+\left(\langle\cos\alpha\rangle_i^{\tau^-}-\langle\cos\alpha\rangle_i^{\tau^+}\right)\,.
\end{align}
As will be shown in the next section (see Eqs.~\eqref{eq:cosalsumi}--\eqref{eq:efun}), the sum $\langle\cos\alpha\rangle_i^{\tau^-} + \langle\cos\alpha\rangle_i^{\tau^+}$ in the first term of Eq.~\eqref{eq:ACPNP} could receive not only from the SM but also from the NP contribution (proportional to the real part of the NP couplings), while the difference $\langle\cos\alpha\rangle_i^{\tau^-} - \langle\cos\alpha\rangle_i^{\tau^+}$ in the second term arises only from the NP part (proportional to the imaginary part of the NP couplings). A global fit of the effective couplings of the non-standard four-fermion operators has been performed under the combined constraints from the branching ratios of $\tau^-\to K_S\pi^-\nu_\tau$, $\tau^-\to K^-\eta\nu_\tau$ and $\tau^-\to K^-\nu_\tau$ decays, as well as the decay spectrum of $\tau^-\to K_S\pi^-\nu_\tau$ decay, with the real parts of the non-standard scalar and tensor couplings given respectively by $\mathrm{Re}[\hat{\epsilon}_S]=(0.8^{+0.8}_{-0.9}\pm0.3)\%$ and $\mathrm{Re}[\hat{\epsilon}_T]=(0.9\pm0.7\pm0.4)\%$ at the scale $\mu_\tau=2$~GeV in $\mathrm{\overline{MS}}$ scheme~\cite{Gonzalez-Solis:2019owk}, being therefore negligible with respect to the SM contribution (which has been normalized to one). This, together with the value of $A^{CP}_K$ given by Eq.~\eqref{eq:ACPK}, implies that the dominant NP impact on the total $CP$ asymmetry $A_i^{CP}$ comes only from the second term of Eq.~\eqref{eq:ACPNP}. We will detail in the next section the NP contributions by performing a model-independent analysis of $\tau^\pm\to K^0(\bar K^0)\pi^\pm\bar\nu_\tau(\nu_\tau)$ decays within a general low-energy EFT framework including four-fermion operators up to dimension-six.

\section{General EFT analysis of \boldmath{$\tau^\pm\to K^0(\bar K^0)\pi^\pm\bar\nu_\tau(\nu_\tau)$ decays}}
\label{sec:Effective}

The $\tau^-\to \bar{K}^0\pi^-\nu_\tau$ decay and its $CP$-conjugated process are mediated at the partonic level by the strangeness-changing $\tau^-\to \nu_\tau \bar{u}s$ and $\tau^+\to \bar\nu_\tau u\bar{s}$ transitions, respectively. Assuming the absence of other light degrees of freedom except for the SM ones below the electroweak scale, as well as the Lorentz and the $SU(3)_C\times U(1)_{\text{em}}$ invariance, the most general low-energy effective Lagrangian governing the strangeness-changing hadronic $\tau$ decays can be written as~\cite{Cirigliano:2009wk,Bhattacharya:2011qm,Miranda:2018cpf,Chang:2014iba,Gonzalez-Alonso:2016etj,Cirigliano:2017tqn,Cirigliano:2018dyk,Garces:2017jpz,Rendon:2019awg,Gonzalez-Solis:2019lze}\footnote{Here we have included neither the right-handed nor the wrong-flavour neutrino fields, which in any case do not interfere with the SM amplitude and thus contribute only at $\mathcal{O}(\epsilon_i^2)$ to the observables considered. One should also keep in mind that, unless some NP is assumed between the characteristic scale $\mu_\tau=2~\mathrm{GeV}$ and the electroweak scale, the low-energy effective Lagrangian given by Eq.~\eqref{eq:Efective_Lagrangian} appears generally as an $SU(2)_L$-invariant form, being part of the SMEFT Lagrangian~\cite{Buchmuller:1985jz,Grzadkowski:2010es,Brivio:2017vri,Alonso:2014csa}. This implies that the effective couplings of the operators contributing to the $\tau\to K_S\pi\nu_\tau$ decays could also be constrained by other processes, such as the neutron EDM and the $D^0-\bar{D}^0$ mixing~\cite{Cirigliano:2017tqn}, which will be detained in subsection~\ref{subsec:otherbounds}.}
\begin{align}\label{eq:Efective_Lagrangian}
\mathcal{L}_\mathrm{eff}=&-\frac{G^0_{F}V_{us}}{\sqrt{2}}\,\Big\{(1+\epsilon_{L})\,\bar{\tau}\gamma_{\mu}
(1-\gamma_{5})\nu_{\tau}\cdot\bar{u}\gamma^{\mu}(1-\gamma_5)s+  \epsilon_{R}\,\bar{\tau}\gamma_{\mu}(1-\gamma_{5})\nu_{\tau}\cdot\bar{u}\gamma^{\mu}(1+\gamma_5)s\,\nonumber\\[0.15cm]
&+\bar{\tau}(1-\gamma_{5})\nu_{\tau}\cdot\bar{u}\left[\epsilon_{S}-\epsilon_{P}\gamma_{5}\right]s+
\epsilon_{T}\,\bar{\tau}\sigma_{\mu\nu}(1-\gamma_{5})\nu_{\tau}\cdot\bar{u}\sigma^{\mu\nu}(1-\gamma_{5})s\Big\}
+\mathrm{h.c.}\,\nonumber\\[0.2cm]
=&-\frac{G^0_{F}V_{us}}{\sqrt{2}}\,(1+\epsilon_{L}+\epsilon_{R})\,\Big\{\bar{\tau}\gamma_{\mu}
(1-\gamma_{5})\nu_{\tau}\cdot\bar{u}\left[\gamma^{\mu}-(1-2\,\hat{\epsilon}_{R})\gamma^{\mu}\gamma_{5}\right]s\,\nonumber\\[0.15cm]
&+\bar{\tau}(1-\gamma_{5})\nu_{\tau}\cdot\bar{u}\left[\hat{\epsilon}_{S}-\hat{\epsilon}_{P}\gamma_{5}\right]s+
2\,\hat{\epsilon}_{T}\,\bar{\tau}\sigma_{\mu\nu}(1-\gamma_{5})\nu_{\tau}\cdot\bar{u}\sigma^{\mu\nu}s\Big\}+
\mathrm{h.c.}\,,
\end{align}
where $G^0_F$ is the Fermi constant in the absence of non-standard NP contributions, and $V_{us}$ is the CKM matrix element involved in the decays. The effective couplings $\epsilon_{i}$ parametrize the non-standard NP contributions and can be generally complex, with the SM case recovered by setting all $\epsilon_{i}=0$. The hatted couplings $\hat{\epsilon}_{i}=\epsilon_{i}/(1+\epsilon_L+\epsilon_R)$ for $i=R, S, P, T$ have also been introduced in the second equation, where the corresponding quark currents possess definite parities and are therefore more convenient to describe the vacuum to $K\pi$ matrix elements due to parity conservation of strong interactions~\cite{Garces:2017jpz,Rendon:2019awg}. It is obvious from Eq.~\eqref{eq:Efective_Lagrangian} that, working at the linear order in $\hat{\epsilon}_{i}$, the decays are insensitive to the non-standard spin-$1$ charged-current interactions because the overall dependence on the combination $\epsilon_{L}+\epsilon_{R}$ cannot be isolated and is generally subsumed in the determination of $G^\mathrm{exp}_F$~\cite{Cirigliano:2009wk,Bhattacharya:2011qm,Miranda:2018cpf}. Throughout the work, we will use the abbreviation $G_F=G^0_F(1+\epsilon_{L}+\epsilon_{R})$, with the superscript `$\mathrm{exp}$' omitted, and quote numerical bounds on $\hat{\epsilon}_{i}$ at the renormalization scale $\mu_\tau=2$~GeV in the $\mathrm{\overline{MS}}$ scheme, unless specified otherwise.

With the most general low-energy effective Lagrangian at hand, we can now perform a model-independent analysis of the $\tau^\pm\to K^0(\bar K^0)\pi^\pm\bar\nu_\tau(\nu_\tau)$ decays. Here we present only the formulae for the $\tau^-\to \bar{K}^0\pi^-\nu_\tau$ decay, while the ones for the $CP$-conjugated process $\tau^+\to K^0\pi^+\bar{\nu}_\tau$ can be easily obtained from the former with the replacements $V_{us}\to V_{us}^\ast$ and $\hat{\epsilon}_i\to\hat{\epsilon}_i^\ast$. Due to parity conservation of strong interactions, we find that only the vector, scalar, and tensor hadronic currents have non-zero contributions to the process, with the resulting decay amplitude given by\footnote{The factor $2$ in front of the tensor-interaction term here as well as in the second equation of Eq.~\eqref{eq:Efective_Lagrangian} results from the Dirac-algebra identity $\sigma^{\mu\nu}\gamma_5=\frac{i}{2}\epsilon^{\mu\nu\alpha\beta}\sigma_{\alpha\beta}$, with the convention $\epsilon^{0123}=-\epsilon_{0123}=1$.}
\begin{equation}
\mathcal{M}(\tau^-\to \bar{K}^0\pi^-\nu_\tau)=\frac{G_FV_{us}}{\sqrt{2}}\,\Big[L_\mu H^\mu+\hat{\epsilon}_S^\ast\,LH+2\,\hat{\epsilon}_T^\ast\, L_{\mu\nu}H^{\mu\nu}\Big]\,,
\end{equation}
where $L_{(i)}$ and $H^{(i)}$ denote the leptonic and the hadronic currents respectively, with
\begin{align}
L&=\bar{u}(p_{\nu_\tau})(1+\gamma_5)u(p_\tau)\,,\\[0.2cm]
L_\mu&=\bar{u}(p_{\nu_\tau})\gamma_\mu(1-\gamma_5)u(p_\tau)\,,\\[0.2cm]
L_{\mu\nu}&=\bar{u}(p_{\nu_\tau})\sigma_{\mu\nu}(1+\gamma_5)u(p_\tau)\,,
\end{align}
and
\begin{align}
H&=\langle \bar{K}^0(p_K)\pi^-(p_\pi)|\bar{s}u|0\rangle=\frac{\Delta_{K\pi}}{m_s-m_u}\,F_0(s)\,,\label{eq:SHad}\\[0.2cm]
H^\mu&=\langle \bar{K}^0(p_K)\pi^-(p_\pi)|\bar{s}\gamma^\mu u|0\rangle=\left[(p_K-p_\pi)^\mu-\frac{\Delta_{K\pi}}{s}\,q^\mu\right]\,F_+(s)+\frac{\Delta_{K\pi}}{s}\,q^\mu\,F_0(s)\,,\label{eq:VHad}\\[0.2cm]
H^{\mu\nu}&=\langle \bar{K}^0(p_K)\pi^-(p_\pi)|\bar{s}\sigma^{\mu\nu}u|0\rangle=iF_T(s)\,\left(p_K^\mu p_\pi^{\nu}-p_\pi^{\mu}p_K^\nu\right)\,.\label{eq:THad}
\end{align}
Here $s=(p_K+p_\pi)^2$, $q^\mu=(p_K+p_\pi)^\mu$, $\Delta_{K\pi}=M_K^2-M_{\pi}^2$, and $F_0(s)$, $F_+(s)$, and $F_T(s)$ stand respectively for the $K\pi$ scalar, vector, and tensor form factors, which encode the hadronization information of the corresponding quark currents between the vacuum and $K\pi$ final state. Note that the equations of motion have been applied to obtain the hadronic matrix element of the scalar current (Eq.~\eqref{eq:SHad}) from that of the vector current (Eq.~\eqref{eq:VHad}). As the form-factor phases fitted via a superposition of Breit-Wigner functions with complex coefficients do not vanish at threshold and violate the Watson's theorem long before the higher resonances start to play an effect~\cite{Cirigliano:2017tqn,Rendon:2019awg,Chen:2019vbr}, we cannot rely on the formalism developed in Refs.~\cite{Epifanov:2007rf,Paramesvaran:2009ec,Finkemeier:1995sr,Finkemeier:1996dh} to study the $CP$ asymmetries in $\tau\to K_S\pi\nu_\tau$ decays. Instead, we will adopt the thrice-subtracted (for the vector form factor)~\cite{Boito:2008fq,Boito:2010me}, the coupled-channel (for the scalar form factor)~\cite{Jamin:2000wn,Jamin:2001zq,Jamin:2006tj}, and the once-subtracted (for the tensor form factor)~\cite{Cirigliano:2017tqn,Rendon:2019awg,Chen:2019vbr}
dispersive representations, which warrant the properties of both unitarity and analyticity, and contain a full knowledge of QCD in both the perturbative and non-perturbative regimes. For convenience, their explicit expressions are collected in the appendix.

Working in the $K\pi$ rest frame and after integrating over the unobserved neutrino direction, we can write the double differential decay width of $\tau^-\to \bar{K}^0\pi^-\nu_\tau$ decay as
\begin{align}\label{eq:dG2dsdcos}
&\frac{d^2\Gamma(\tau^-\to \bar{K}^0\pi^-\nu_\tau)}{ds\,d\cos\alpha}=\frac{G_{F}^{2}|V_{us}|^{2} m_{\tau}^{3} S_\mathrm{EW}}{512 \pi^{3} s^3}\,\left(1-\frac{s}{m_{\tau}^{2}}\right)^{2} \lambda^{\frac{1}{2}}\left(s, M_{K}^{2}, M_{\pi}^{2}\right)\nonumber\\[0.15cm]
&\times\Bigg\lbrace\lambda\left(s, M_{K}^{2}, M_{\pi}^{2}\right)\,\left[\frac{s}{m_\tau^2}\!+\!\left(1\!-\!\frac{s}{m_\tau^2}\right)\cos^2\alpha\right]|F_+(s)|^2+\Delta_{K\pi}^2\left|1\!+\!\frac{\hat{\epsilon}_S\,s}{m_\tau(m_s\!-\!m_u)}\right|^2|F_0(s)|^2\nonumber\\[0.15cm]
&\hspace{0.2cm}+4\lambda\left(s, M_{K}^{2}, M_{\pi}^{2}\right)\,\left[s|\hat{\epsilon}_T|^2\left(1\!-\!\left(1\!-\!\frac{s}{m_\tau^2}\right)\cos^2\alpha\right)|F_T(s)|^2-\frac{s}{m_\tau}\mathrm{Re}\left[\hat{\epsilon}_T F_+(s)F_T^\ast(s)\right]\right]\nonumber\\[0.15cm]
&\hspace{0.2cm}-2\Delta_{K\pi}\,\lambda^{\frac{1}{2}}\left(s, M_{K}^{2}, M_{\pi}^{2}\right)\,\mathrm{Re}\left[\left(1+\frac{\hat{\epsilon}_S\,s}{m_\tau(m_s-m_u)}\right)F_+(s)F_0^\ast(s)\right]\cos\alpha\nonumber\\[0.15cm]
&\hspace{0.2cm}+\frac{4s}{m_\tau}\,\Delta_{K\pi}\,\lambda^{\frac{1}{2}}\left(s, M_{K}^{2}, M_{\pi}^{2}\right)\,\mathrm{Re}\left[\hat{\epsilon}^\ast_T\left(1+\frac{\hat{\epsilon}_S\,s}{m_\tau(m_s-m_u)}\right)F_T(s)F_0^\ast(s)\right]\cos\alpha
\Bigg\rbrace\,,
\end{align}
where $\lambda\left(s,M_K^2,M_{\pi}^2\right)=\left[s-(M_{K}+M_{\pi})^2\right]\left[s-(M_{K}-M_{\pi})^2\right]$ is the usual K{\"a}ll{\'e}n function, and $S_\mathrm{EW}=1.0201(3)$ encodes the short-distance electroweak radiative correction~\cite{Erler:2002mv,Marciano:1985pd,Marciano:1988vm,Braaten:1990ef}, which has been simply written as an overall factor, although it affects only the SM contribution~\cite{Rendon:2019awg}. Integrating Eq.~\eqref{eq:dG2dsdcos} over $\cos\alpha$, one then arrives at the differential decay width as a function of the $K\pi$ invariant mass squared $s$, which is given explicitly by
\begin{align}\label{eq:dG2ds}
&\frac{d\Gamma(\tau^-\to \bar{K}^0\pi^-\nu_\tau)}{ds}=\frac{G_{F}^{2}|V_{u s}|^{2} m_{\tau}^{3} S_\mathrm{EW}}{512 \pi^{3} s^3}\,\left(1-\frac{s}{m_{\tau}^{2}}\right)^{2}\,\lambda^{\frac{1}{2}}\left(s, M_{K}^{2}, M_{\pi}^{2}\right)\nonumber\\[0.15cm]
&\times\Bigg\lbrace\frac{2}{3}\lambda\left(s, M_{K}^{2}, M_{\pi}^{2}\right)\,\left(1+\frac{2s}{m_\tau^2}\right)|F_+(s)|^2+2\Delta_{K\pi}^2\,\left|1+\frac{\hat{\epsilon}_S\,s}{m_\tau(m_s-m_u)}\right|^2|F_0(s)|^2\nonumber\\[0.15cm]
&\hspace{0.35cm}+\frac{8}{3}\lambda\left(s, M_{K}^{2}, M_{\pi}^{2}\right)\,\left(s|\hat{\epsilon}_T|^2\left(2+\frac{s}{m_\tau^2}\right)|F_T(s)|^2-\frac{3s}{m_\tau}\,\mathrm{Re}\left[\hat{\epsilon}_T F_+(s)F_T^\ast(s)\right]\right)\Bigg\rbrace\,.
\end{align}
Integrating further Eq.~\eqref{eq:dG2ds} over $s$, one finally obtains the decay width of $\tau^-\to\bar{K}^0\pi^-\nu_\tau$, which includes both the SM and the NP contributions. 

In order to generate a non-vanishing direct CPV in the $\tau\to K_S\pi\nu_\tau$ decay rates, we need at least two amplitudes with different weak and strong phases, which implies that the terms proportional to the modulus squared of each form factor in Eq.~\eqref{eq:dG2ds} have no contributions to the decay-rate asymmetry, leaving therefore the vector-tensor interference (the last term in the curly bracket of Eq.~\eqref{eq:dG2ds}) as the only possible mechanism~\cite{Devi:2013gya}. The resulting $CP$ asymmetry in the decay rates can then be written as~\cite{Cirigliano:2017tqn,Rendon:2019awg,Chen:2019vbr}
\begin{align}\label{eq:intACP}
A_{CP}^\mathrm{rate}(\tau\to K\pi\nu_\tau)=&\frac{\Gamma(\tau^+\to K^0\pi^+\bar{\nu}_\tau)-\Gamma(\tau^-\to \bar{K}^0\pi^-\nu_\tau)}{\Gamma(\tau^+\to K^0\pi^+\bar{\nu}_\tau)+\Gamma(\tau^-\to \bar{K}^0\pi^-\nu_\tau)}\nonumber\\[0.2cm]
=&\frac{\mathrm{Im}[\hat{\epsilon}_{T}]\,G_{F}^{2}|V_{u s}|^{2} S_\mathrm{EW}}{128\,\pi^{3}\, m_{\tau}^{2}\, \Gamma(\tau \to K_S\pi\nu_{\tau})}\,\int_{s_{K\pi }}^{m_{\tau}^{2}} d s\left(1-\frac{m_{\tau}^{2}}{s}\right)^{2}\,\lambda^{\frac{3}{2}}\left(s, M_{K}^{2}, M_{\pi}^{2}\right)\nonumber\\[0.15cm]
&\hspace{3.5cm}\times|F_{T}(s)|\,|F_{+}(s)|\,\sin\left[\delta_T(s)-\delta_+(s)\right]\,,
\end{align}
where $s_{K\pi}=(M_K+M_{\pi})^2$ denotes the threshold of the $K\pi$ invariant mass squared $s$, and $\delta_{+}(s)$ and $\delta_{T}(s)$ stand for the phases of the $K\pi$ vector and tensor form factors, respectively. The scenario with new tensor interactions was firstly proposed in Ref.~\cite{Devi:2013gya} to accommodate the $2.8~\sigma$ tension observed between the SM prediction~\cite{Bigi:2005ts,Calderon:2007rg,Grossman:2011zk,Chen:2019vbr} and the BaBar measurement~\cite{BABAR:2011aa} of the decay-rate asymmetry in $\tau\to K_S\pi\nu_\tau$ decays. However, as mentioned already in section~\ref{sec:intro}, such a scenario has been claimed to be infeasible due to the strong suppression of the relative phase between the $K\pi$ vector and tensor form factors as well as the combined constraints from other observables~\cite{Cirigliano:2017tqn,Rendon:2019awg,Chen:2019vbr}.

Another interesting observable in $\tau\to K_S\pi\nu_\tau$ decays is the $CP$ asymmetry in the angular distributions, $A_i^{CP}$, defined by Eq.~\eqref{eq:ACPNP}, which shows a strong sensitivity to the non-standard scalar and tensor interactions present in Eq.~\eqref{eq:Efective_Lagrangian}. To obtain $A_i^{CP}$, we must firstly determine the angular observable $\langle\cos\alpha\rangle^{\tau^{-}}(s)$, which is defined as the differential decay width weighted by $\cos\alpha$:
\begin{align}\label{eq:averagecosalpha}
\langle\cos\alpha\rangle^{\tau^{-}}(s)&= \frac{\int_{-1}^{1} \cos \alpha \left(\frac{d^2\Gamma(\tau^-\to \bar{K}^0\pi^-\nu_\tau)}{ds\,d\cos\alpha}\right) d\cos \alpha}{\int_{-1}^{1} \left(\frac{d^2 \Gamma(\tau^-\to \bar{K}^0\pi^-\nu_\tau)}{ds\,d\cos\alpha}\right) d\cos\alpha}\nonumber\\[0.2cm]
&=\frac{N(s)}{D(s)}\,,
\end{align}
with
\begin{align}
N(s)=&-\frac{4}{3}\Delta_{K\pi}\,\lambda^\frac{1}{2}\left(s, M_{K}^{2}, M_{\pi}^{2}\right)\,\mathrm{Re}\left[\left(1+\frac{\hat{\epsilon}_S\,s}{m_\tau(m_s-m_u)}\right)F_+(s)F_0^\ast(s)\right]\nonumber\\[0.15cm]
&+\frac{8s}{3m_\tau}\,\Delta_{K\pi}\,\lambda^\frac{1}{2}\left(s, M_{K}^{2}, M_{\pi}^{2}\right)\,\mathrm{Re}\left[\hat{\epsilon}^\ast_T\left(1+\frac{\hat{\epsilon}_S\,s}{m_\tau(m_s-m_u)}\right)F_T(s)F_0^\ast(s)\right]\,,\label{eq:num}\\[0.2cm]
D(s)=&\frac{2}{3}\lambda\left(s, M_{K}^{2}, M_{\pi}^{2}\right)\,\left(1+\frac{2s}{m_\tau^2}\right)\,|F_+(s)|^2+2\Delta_{K\pi}^2\,\left|1+\frac{\hat{\epsilon}_S\,s}{m_\tau(m_s-m_u)}\right|^2|F_0(s)|^2\nonumber\\[0.15cm]
&+\frac{8}{3}\lambda\left(s, M_{K}^{2}, M_{\pi}^{2}\right)\,\left[s|\hat{\epsilon}_T|^2\left(2+\frac{s}{m_\tau^2}\right)|F_T(s)|^2-\frac{3s}{m_\tau}\mathrm{Re}\left[\hat{\epsilon}_T F_+(s)F_T^\ast(s)\right]\right]\,.\label{eq:den}
\end{align}
It is worth to note that $\langle\cos\alpha\rangle^{\tau^{-}}(s)$ is connected to the forward-backward asymmetry $A_\mathrm{FB}^{\tau^{-}}(s)$ via the relation $\langle\cos\alpha\rangle^{\tau^{-}}(s)=2/3\,A_\mathrm{FB}^{\tau^{-}}(s)$, with the latter defined by~\cite{Beldjoudi:1994hi,Kimura:2014wsa,Gao:2012su}
\begin{align}\label{eq:AFB}
A_\text{FB}^{\tau^-}(s)&=\frac{\int_0^1\frac{d^2\Gamma(\tau^-\to \bar{K}^0\pi^-\nu_\tau)}{ds\,d\cos\alpha}\,d\cos\alpha-\int_{-1}^0\frac{d^2\Gamma(\tau^-\to \bar{K}^0\pi^-\nu_\tau)}{ds\,d\cos\alpha}\,d\cos\alpha}{\int_0^1\frac{d^2\Gamma(\tau^-\to \bar{K}^0\pi^-\nu_\tau)}{ds\,d\cos\alpha}\,d\cos\alpha+\int_{-1}^0\frac{d^2\Gamma(\tau^-\to \bar{K}^0\pi^-\nu_\tau)}{ds\,d\cos\alpha}\,d\cos\alpha}\,.
\end{align}
One can see from Eqs.~\eqref{eq:averagecosalpha}--\eqref{eq:den} that the angular observable $\langle\cos\alpha\rangle^{\tau^{-}}(s)$ (or equivalently the forward-backward asymmetry $A_\text{FB}^{\tau^-}(s)$) is proportional to the factor $\Delta_{K\pi}=M_K^2-M_{\pi}^2$, implying that the observable may also allow us to depict the $SU(3)$-symmetry breaking effect in the decays considered~\cite{Beldjoudi:1994hi}. In addition, it is clear from Eq.~\eqref{eq:num} that there exist two terms contributing to $\langle\cos\alpha\rangle^{\tau^{-}}(s)$, with the first and the second one corresponding to the scalar-vector and the tensor-scalar interference respectively, whereas only the first term remains within the SM obtained after setting $\hat{\epsilon}_S=\hat{\epsilon}_T=0$. 

Experimentally, on the other hand, both the differential decay width $\frac{d\Gamma(\tau^-\to \bar{K}^0\pi^-\nu_\tau)}{ds}$ and the angular observable $\langle\cos\alpha\rangle^{\tau^{-}}(s)$ are usually measured in different bins of the $K\pi$ invariant mass squared $s$. Thus, we can make these observables bin-dependent, such as the observable $\langle\cos \alpha\rangle^{\tau^{-}}_i$ defined by
\begin{align}\label{eq:cosbin}
\langle\cos \alpha\rangle^{\tau^{-}}_i =& \frac{\int_{s_{1,i}}^{s_{2,i}}\int_{-1}^{1} \cos \alpha \left(\frac{d^2\Gamma(\tau^-\to \bar{K}^0\pi^-\nu_\tau)}{ds\,d\cos\alpha}\right)ds\,d\cos \alpha}{\int_{s_{1,i}}^{s_{2,i}}\int_{-1}^{1} \left(\frac{d^2\Gamma(\tau^-\to \bar{K}^0\pi^-\nu_\tau)}{ds\,d\cos\alpha}\right) ds\,d\cos\alpha}\,.
\end{align}
The explicit expression of the $CP$-conjugated observable $\langle\cos \alpha\rangle^{\tau^{+}}_i$ can be easily obtained from Eq.~\eqref{eq:cosbin} with the simple replacements $V_{us}\to V_{us}^\ast$ and $\hat{\epsilon}_i\to\hat{\epsilon}_i^\ast~(i=S,T)$. According to the definitions of Eqs.~\eqref{eq:cossum} and~\eqref{eq:cosdif}, the sum $\langle\cos\alpha\rangle_i^{\tau^-} + \langle\cos\alpha\rangle_i^{\tau^+}$ and the difference $\langle\cos\alpha\rangle_i^{\tau^-} - \langle\cos\alpha\rangle_i^{\tau^+}$ are then given, respectively, by
\begin{align}
\langle\cos \alpha\rangle^{\tau^{-}}_i+\langle\cos \alpha\rangle^{\tau^{+}}_i=\frac{\int_{s_{1,i}}^{s_{2,i}}\, A(s)\,ds}{\int_{s_{1,i}}^{s_{2,i}}\, E(s)\,ds}\,,\label{eq:cosalsumi}\\[0.2cm]
\langle\cos \alpha\rangle^{\tau^{-}}_i-\langle\cos \alpha\rangle^{\tau^{+}}_i=\frac{\int_{s_{1,i}}^{s_{2,i}}\, B(s)\,ds}{\int_{s_{1,i}}^{s_{2,i}}\, E(s)\,ds}\,,\label{eq:cosaldifi}
\end{align}
with
\begin{align}
A(s)=&\frac{G_{F}^{2}|V_{u s}|^{2} m_{\tau}^{3} S_\mathrm{EW}}{512 \pi^{3} s^3}\,\left(1-\frac{s}{m_{\tau}^{2}}\right)^{2}\, \lambda\left(s, M_{K}^{2}, M_{\pi}^{2}\right)\,\Delta_{K\pi}\nonumber\\[0.15cm]
&\hspace{-0.8cm}\times\Bigg\lbrace-\frac{8}{3}\left(1+\frac{\mathrm{Re}[\hat{\epsilon}_S]\,s}{m_\tau(m_s-m_u)}\right)\mathrm{Re}[F_+(s)F^\ast_0(s)]+\frac{16s}{3m_\tau}\mathrm{Re}[\hat{\epsilon}_T]\,\mathrm{Re}[F_T(s)F^\ast_0(s)]\Bigg\rbrace\,,\label{eq:afun}\\[0.2cm]
B(s)=&\frac{G_{F}^{2}|V_{u s}|^{2} m_{\tau}^{3} S_\mathrm{EW}}{512 \pi^{3} s^3}\,\left(1-\frac{s}{m_{\tau}^{2}}\right)^{2}\,\lambda\left(s, M_{K}^{2}, M_{\pi}^{2}\right)\,\Delta_{K\pi}\nonumber\\[0.15cm]
&\hspace{-0.8cm}\times\Bigg\lbrace\frac{8}{3}\frac{\mathrm{Im}[\hat{\epsilon}_S]\,s}{m_\tau(m_s-m_u)} \,\mathrm{Im}[F_+(s)F^\ast_0(s)]+\frac{16s}{3m_\tau}\mathrm{Im}[\hat{\epsilon}_T]\,\mathrm{Im}[F_T(s)F^\ast_0(s)]\Bigg\rbrace\,,\label{eq:bfun}\\[0.2cm]
E(s)=&\frac{G_{F}^{2}|V_{u s}|^{2} m_{\tau}^{3} S_\mathrm{EW}}{512 \pi^{3} s^3}\,\left(1-\frac{s}{m_{\tau}^{2}}\right)^{2}\,\lambda^{\frac{1}{2}}\left(s, M_{K}^{2}, M_{\pi}^{2}\right)\nonumber\\[0.15cm]
&\hspace{-0.8cm}\times\Bigg\lbrace \frac{2}{3}\lambda\left(s, M_{K}^{2}, M_{\pi}^{2}\right)\,\left(1+\frac{2s}{m_\tau^2}\right)|F_+(s)|^2+2\Delta_{K\pi}^2\,\left|1+\frac{\hat{\epsilon}_S\,s}{m_\tau(m_s-m_u)}\right|^2|F_0(s)|^2\nonumber\\[0.15cm]
&\hspace{-0.8cm}+\frac{8}{3}\lambda\left(s, M_{K}^{2}, M_{\pi}^{2}\right)\,\left(s|\hat{\epsilon}_T|^2\left(2+\frac{s}{m_\tau^2}\right)|F_T(s)|^2-\frac{3s}{m_\tau}\mathrm{Re}\,\left[\hat{\epsilon}_T\right]|F_T(s)||F_+(s)|\right)\Bigg\rbrace\,.\label{eq:efun}
\end{align}
Here we have neglected the terms quadratic in the NP parameters to obtain Eqs.~\eqref{eq:afun} and \eqref{eq:bfun}, and taken the approximation $\delta_T(s)\approx\delta_+(s)$ to obtain the last term in the third line of Eq.~\eqref{eq:efun}, since the two phases are strictly equal in the elastic region and deviate slightly from each other only in the inelastic region of the $K\pi$ re-scattering. This feature can be seen clearly from Figure~\ref{fig:FFs} given in the appendix (cf. also Figure~5 of Ref.~\cite{Chen:2019vbr}). 

Substituting Eqs.~\eqref{eq:cosalsumi} and \eqref{eq:cosaldifi} back into Eq.~\eqref{eq:ACPNP}, one obtains immediately the full expression of the $CP$ asymmetry in the angular distributions of $\tau\to K_S\pi\nu_\tau$ decays, which is now given in terms of the effective couplings of the non-standard scalar and tensor interactions. Combining Eqs.~\eqref{eq:ACPNP} and \eqref{eq:dG2ds}, one can then obtain bounds on the NP parameters under the combined constraints from the available data on the $CP$ asymmetries measured in four $K\pi$ invariant-mass bins by Belle~\cite{Bischofberger:2011pw} as well as the branching ratio of $\tau^-\to K_S\pi^-\nu_\tau$ decay~\cite{Epifanov:2007rf}. This will be explored in the next section.

\section{Numerical results and discussions}
\label{sec:numerical}

\subsection{Input parameters and choice of the NP couplings}

Before presenting our numerical results, we firstly collect in Table~\ref{tab:input} all the input parameters used throughout this work; for any further details, the readers are referred to the references therein. For the $K\pi$ scalar form factor, we adopt the numerical results obtained from a combined analysis of the $\tau^{-}\to K_{S}\pi^{-}\nu_{\tau}$ and $\tau^{-}\to K^{-}\eta\nu_{\tau}$ decays~\cite{Escribano:2014joa}, based on the coupled-channel dispersive representation~\cite{Jamin:2000wn,Jamin:2001zq,Jamin:2006tj}.\footnote{We thank Pablo Roig for providing us with the necessary numerical tables obtained in Ref.~\cite{Escribano:2014joa}.} Detailed information of the $K\pi$ tensor form factor can be found in the appendix. For each observable, the experimental error is obtained by simply adding the statistical and systematic ones in quadrature, while the theoretical uncertainty is calculated by varying each input parameter within its corresponding range and then adding the resulting individual uncertainty in quadrature.

\begin{table}[t]
	\tabcolsep 0.068in
	\renewcommand\arraystretch{1.43}
	\begin{center}
		\caption{\small Summary of the input parameters used throughout this work. The up- and strange-quark masses are given in the $\mathrm{\overline{MS}}$ scheme at the scale $\mu=2~\mathrm{GeV}$~\cite{Zyla:2020zbs}. The hadronic matrix elements of $\Delta C = 2$ four-fermion operators contributing to the $D^0-\bar{D}^0$ mixing are evaluated in the $\mathrm{\overline{MS}}$-NDR scheme at the scale $\mu=3~\mathrm{GeV}$~\cite{Bazavov:2017weg}. \label{tab:input} } 
		\vspace{0.2cm}
		\begin{tabular}{|c|c|c|c|c|}
			\hline\hline
			\multicolumn{5}{|l|}{QCD and electroweak parameters}\\
			\hline
			\multicolumn{2}{|c|}{$G_F[\text{GeV}^{-2}]$~\cite{Zyla:2020zbs}} & $\left|V_{us}F_+(0)\right|$~\cite{Moulson:2017ive} & $F_{\pi}~[\text{MeV}]$~\cite{Zyla:2020zbs} & $F_K~[\text{MeV}]$~\cite{Zyla:2020zbs}\\
			\hline
			\multicolumn{2}{|c|}{$1.1663787(6)\times10^{-5}$} & $0.21654(41)$ & $92.3(1)$  & $1.198F_\pi$\\
			\hline
			\multicolumn{5}{|l|}{Particle as well as the up- and strange-quark masses~\cite{Zyla:2020zbs}}\\
			\hline
			$m_{\tau}~[\text{MeV}]$ & $M_{K^0}~[\text{MeV}]$ & $M_{\pi^-}~[\text{MeV}]$ & $m_u~[\text{MeV}]$ & $m_s~[\text{MeV}]$ \\
			\hline
			$1776.86$ & $497.61$ & $139.57$ & $2.16^{+0.49}_{-0.26}$ & $93^{+11}_{-\phantom{0}5}$ \\
			\hline
			\multicolumn{5}{|l|}{Parameters in the $K\pi$ vector form factor with $s_{cut}=4~\text{GeV}^2$~\cite{Boito:2008fq}}\\
			\hline
			$m_{K^*}~[\text{MeV}]$ & $\gamma_{K^*}~[\text{MeV}]$ & $m_{K^{*\prime}}~[\text{MeV}]$ & $\gamma_{K^{*\prime}}~[\text{MeV}]$& $\gamma$\\
			\hline
			$943.41\pm0.59$& $66.72\pm0.87$ & $1374\pm45$ & $240\pm131$ & $-0.039\pm0.020$\\
			\hline
			$M_{K^*}~[\text{MeV}]$ &\multicolumn{2}{c|}{$\lambda_+^{\prime}$} & \multicolumn{2}{c|}{$\lambda_+^{\prime\prime}$ }\\
			\hline
			$892.01\pm0.92$&\multicolumn{2}{c|}{$(24.66\pm0.77)\times10^{-3}$} & \multicolumn{2}{c|}{$(11.99\pm0.20)\times10^{-4}$}\\         		
			\hline
			\multicolumn{5}{|l|}{CKM matrix elements~\cite{Zyla:2020zbs}}\\
			\hline
			$V_{ud}$ & $V_{us}$ & $V_{cd}$ & \multicolumn{2}{c|}{$V_{cs}$} \\
			\hline
			$0.97370(14)$ & $0.2245(8)$ & $-0.221(4)$ & \multicolumn{2}{c|}{$0.987(11)$}\\        		
			\hline
			\multicolumn{5}{|l|}{$CP$-violating parameters in the neutral kaon system~\cite{Zyla:2020zbs}}\\
			\hline
			$\left|\eta_{+-}\right|\times10^{3}$ & \multicolumn{2}{c|}{$\phi_{+-}$} & \multicolumn{2}{c|}{$\mathrm{Re}(\epsilon_K)\times10^{3}$}\\
			\hline
			$2.232\pm0.011$ & \multicolumn{2}{c|}{$(43.51\pm0.05)^{\circ}$} & \multicolumn{2}{c|}{$1.66\pm0.02$}\\
			\hline
		    \multicolumn{5}{|l|}{Other input parameters}\\
			\hline
            \multicolumn{2}{|c|}{$\mathcal{B}(\tau^-\to K_S\pi^-\nu_\tau)$~\cite{Epifanov:2007rf}} & $\tau$ lifetime~\cite{Zyla:2020zbs} & $\mathrm{Re}[\hat{\epsilon}_S]$~\cite{Gonzalez-Solis:2019owk} & $\mathrm{Re}[\hat{\epsilon}_T]$~\cite{Gonzalez-Solis:2019owk} \\
			\hline
			\multicolumn{2}{|c|}{$(4.04\pm0.02\pm 0.13)\times10^{-3}$} & $290.3\times10^{-15}~\text{s}$ & $(0.8^{+0.8}_{-0.9}\pm0.3)\%$ & $(0.9\pm0.7\pm0.4)\%$\\
			\hline
			\multicolumn{2}{|c|}{$|d_n|~[e\,\mathrm{cm}]$~\cite{nEDM:2020crw}} & $g_T^u(2~{\rm GeV})$~\cite{Gupta:2018lvp} & $M_{D^0}~[\rm{MeV}]$~\cite{Zyla:2020zbs} & $D^0$ lifetime~\cite{Zyla:2020zbs}\\
			\hline
			\multicolumn{2}{|c|}{$<1.8\times10^{-26}$} & $-0.204(11)(10)$ & $1864.83(5)$ & $410.1(1.5)\times10^{-15}~\text{s}$\\
			\hline
			\multicolumn{2}{|c|}{$x_{12}$~\cite{Amhis:2019ckw}} & $\phi_{12}$~\cite{Amhis:2019ckw} & \multicolumn{2}{c|}{$\langle D^0|(\bar{c}_L^{\alpha} u_R^{\alpha(\beta)})(\bar{c}_L^{\beta} u_R^{\beta(\alpha)})|\bar{D}^0\rangle~(3~\text{GeV})$~\cite{Bazavov:2017weg}}\\
			\hline
			\multicolumn{2}{|c|}{$(0.409\pm0.048)\%$} & $\left(0.58^{+0.91}_{-0.90}\right)^{\circ}$ & \multicolumn{2}{c|}{$-0.1561(70)(31)~[0.0464(31)(9)]~\text{GeV}^4$}\\
			\hline \hline
		\end{tabular}
	\end{center}
\vspace*{-0.4cm}
\end{table}

To generate a non-vanishing $CP$ asymmetry in the angular distributions of $\tau\to K_S\pi\nu_\tau$ decays, one usually resorts to a charged-scalar exchange with complex couplings, the contribution of which can then interfere with the SM amplitude~\cite{Tsai:1996ps,Kuhn:1996dv,Kimura:2009pm,Kimura:2014wsa}. The $CP$ asymmetry induced by such a possibility has been searched for by the CLEO~\cite{Bonvicini:2001xz} and Belle~\cite{Bischofberger:2011pw} collaborations, by measuring the difference between the angular distributions of $\tau^{+}$ and $\tau^{-}$ decays. Especially, the $CP$ asymmetries measured in four bins of the $K\pi$ invariant mass $\sqrt{s}$ are all found to be compatible with zero with a precision of $\mathcal{O}(10^{-3})$, except for a $1.9\sigma$ deviation for the lowest bin~\cite{Bischofberger:2011pw}. The resulting limit on the $CP$-violating parameter $\mathrm{Im}(\eta_S)$ (which is equivalent to $-\mathrm{Im}[\hat{\epsilon}_S]$ of this work) at the $90\%$ confidence level (C.L.) is in the range $|\mathrm{Im}(\eta_S)|<0.026$ or better, depending on the parametrization used to describe the hadronic form factors~\cite{Bischofberger:2011pw}, and improves upon the previous CLEO limit~\cite{Bonvicini:2001xz} by about one order of magnitude. This demonstrates the great potential of the $CP$-violating angular observables in probing the non-standard scalar interactions.  

In this work, we will update the constraints on the NP parameters by using the same Belle data set~\cite{Bischofberger:2011pw},\footnote{Here we will not use in our parameter constraints the CLEO data~\cite{Bonvicini:2001xz}, because the Belle limit on the parameter $\mathrm{Im}(\eta_S)$ is about one order of magnitude stronger than that obtained by CLEO.} as well as the branching ratio of $\tau^-\to K_S\pi^-\nu_\tau$ decay~\cite{Epifanov:2007rf}, with the following improvements. Firstly, the whole analysis will be performed in a general model-independent framework including both the non-standard scalar and tensor interactions. Secondly, the more reliable dispersive representations rather than the Breit-Wigner parametrizations of the $K\pi$ form factors will be used throughout this work. Thirdly, the non-zero $CP$ asymmetry in the angular distributions induced by the well-established CPV in $K^0-\bar{K}^0$ mixing, as pointed out for the first time by three of us in Ref.~\cite{Chen:2020uxi}, will also be taken into account during the analysis. Finally, as argued already in section~\ref{sec:CPV}, once the combined constraints from the branching ratios of $\tau^-\to K_S\pi^-\nu_\tau$, $\tau^-\to K^-\eta\nu_\tau$ and $\tau^-\to K^-\nu_\tau$ decays, as well as the decay spectrum of $\tau^-\to K_S\pi^-\nu_\tau$ decay are taken into account~\cite{Gonzalez-Solis:2019owk}, the impact of the real parts of the non-standard scalar and tensor couplings, $\mathrm{Re}[\hat{\epsilon}_S]$ and $\mathrm{Re}[\hat{\epsilon}_T]$, on the $CP$ asymmetry in the angular distributions of $\tau\to K_S\pi\nu_\tau$ decays will be negligible. This, in turn, implies that the Belle data on the $CP$ asymmetries measured in four $K\pi$ invariant-mass bins do not contribute significantly to the bounds on these two parameters. Thus, we will simply take the more stringent bounds on $\mathrm{Re}[\hat{\epsilon}_S]$ and $\mathrm{Re}[\hat{\epsilon}_T]$ obtained in Ref.~\cite{Gonzalez-Solis:2019owk}, and are therefore left with two NP parameters to be constrained, \textit{i.e.}, $\mathrm{Im}[\hat{\epsilon}_S]$ and $\mathrm{Im}[\hat{\epsilon}_T]$, denoting the imaginary parts of the non-standard scalar and tensor couplings, respectively.

\subsection{Updated SM predictions including the detector efficiencies}

So far, the formulae derived in section~\ref{sec:Effective} for the $CP$-violating angular observable $A_{i}^{CP}$ follow the convention used in Ref.~\cite{Kou:2018nap}. In order to make full use of the Belle data to constrain the NP parameters, we need firstly rewrite the observable $A_{i}^{CP}$ in terms of the angular variables adopted by Belle~\cite{Bischofberger:2011pw}. In addition, the specific experimental conditions always play an important role in getting the measured results, and their effects, which can be parametrized as the detector efficiencies, must be taken into account properly. For this purpose, we will derive an expression of $A_{i}^{CP}$ that connects directly the theoretical prediction with the Belle measurement~\cite{Bischofberger:2011pw}. 

\begin{figure}[ht]
	\centering
	\includegraphics[width=0.55\textwidth]{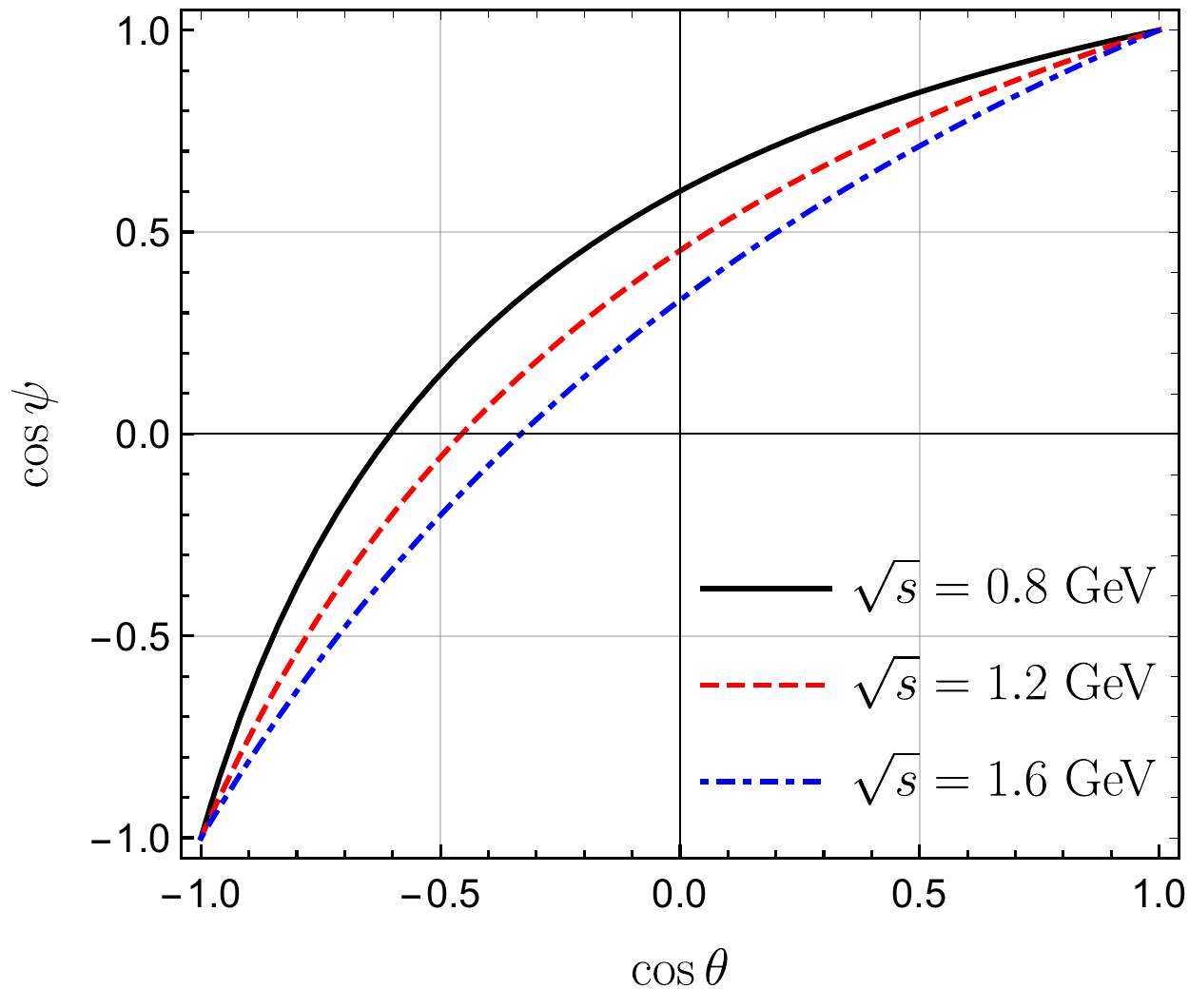}
	\caption{\small Numerical relation between $\cos\theta$ and $\cos\psi$ for three fixed values of $\sqrt{s}$, with $\sqrt{s}=0.8$ (black solid line), $1.2$ (red dashed line), and $1.6$~GeV (blue dot-dashed line), respectively. \label{fig:cos} }
\end{figure}

The original formula used to extract the $CP$-violating parameter $\mathrm{Im}(\eta_S)$ from the measured $CP$ asymmetries is given approximately by~\cite{Bischofberger:2011pw}
\begin{align}\label{eq:Acpi}
A_i^{CP}\simeq\mathrm{Im}(\eta_S)\,\frac{N_s}{n_i}\,\int_{s_{1,i}}^{s_{2,i}}C(s)\,\frac{\mathrm{Im}[F_+(s)F_H^\ast(s)]}{m_\tau}\,ds\,,
\end{align}
where $n_i$ is the observed number of $\tau\to K_S\pi\nu_\tau$ events in the $i$-th bin ($s\in[s_{1,i}, s_{2,i}]$), and $N_s=\sum_i n_i$ denotes the total number of $\tau\to K_S\pi\nu_\tau$ events observed~\cite{Bischofberger:2011pw}. The form factor $F_H(s)$ encodes the hadronization of the scalar quark current, and is related to the scalar form factor $F_0(s)$ defined by Eq.~\eqref{eq:SHad} via the relation $F_H(s)=\frac{\Delta_{K\pi}}{m_s-m_u}F_0(s)$. The function $C(s)$ accounts for the detector efficiencies (including both the total efficiency $\epsilon_\mathrm{tot}$ and the three-dimensional detector efficiency $\epsilon(s,\cos\beta,\cos\psi)$) as well as all the model-independent terms, and is obtained after numerical integration over $\cos\theta$ and $\cos\beta$~\cite{Bischofberger:2011pw}:
\begin{align}\label{eq:Cfunc}
 C(s)=&-\frac{1}{\Gamma(\tau\to K_S\pi\nu_\tau)}\,\frac{G_F^2}{2m_\tau}\,|V_{us}|^2\,\frac{1}{(4\pi)^3}\,\frac{(m_\tau^2-s)^2}{s}\,\frac{\lambda(s,M_K^2,M_\pi^2)}{4s}\nonumber\\[0.2cm] &\times\int\int\frac{\epsilon(s,\cos\beta,\cos\psi)}{\epsilon_\mathrm{tot}}\,\cos^2\beta\,\cos^2\psi\,d\cos\theta\,d\cos\beta\,.
\end{align}
Here the variable $\theta$ is defined as the angle between the direction opposite to that of the $e^+e^-$ CM system and the direction of the hadronic system in the $\tau$ rest frame and, as mentioned already in section~\ref{sec:intro}, is used only by Belle~\cite{Bischofberger:2011pw} but has not been adopted in deriving the formulae of section~\ref{sec:Effective}, where the variable $\psi$ defined as the angle between the direction of the $e^+e^-$ CM frame and the direction of the $\tau$ as seen from the hadronic rest frame is used instead. Thus, we must firstly find out the relation between these two variables. This can be easily achieved, since we have the relation $\cos\alpha=\cos\beta\cos\psi$ (which is obtained after integrating over the azimuthal angle $\phi$~\cite{Kou:2018nap}), and both $\cos\theta$ and $\cos\psi$ can be calculated from the hadronic energy $E_h$ measured in the $e^+e^-$ CM system~\cite{Kuhn:1982di,Kuhn:1991cc,Kuhn:1992nz,Kuhn:1996dv}:
\begin{align}
 \cos\theta&=\frac{2x m_\tau^2-m_\tau^2-s}{(m_\tau^2-s)\,\sqrt{1-m_\tau^2/E^2_\mathrm{beam}}}\,,\label{eq:costheta}\\[0.2cm]
 \cos\psi&=\frac{x(m_\tau^2+s)-2s}{(m_\tau^2-s)\,\sqrt{x^2-s/E^2_\mathrm{beam}}}\,,\label{eq:cospsi}
\end{align}
where $x=E_h/E_\mathrm{beam}$, and $E_\mathrm{beam}=10.58~\mathrm{GeV}$ is the CM energy of Belle~\cite{Bischofberger:2011pw}. One can firstly solve $x$ from Eq.~\eqref{eq:costheta}, and then substitute it into Eq.~\eqref{eq:cospsi} to obtain $\cos\psi$ as a function of $\cos\theta$. The resulting numerical relation between $\cos\theta$ and $\cos\psi$ is shown in Figure~\ref{fig:cos} for three fixed values of the $K\pi$ invariant mass, with $\sqrt{s}=0.8$ (black solid line), $1.2$ (red dashed line), and $1.6$~GeV (blue dot-dashed line), respectively.

Notice that in Ref.~\cite{Bischofberger:2011pw} a more convenient parametrization form of the function $C(s)$ is given as a seventh-order polynomial,
\begin{align}\label{eq:polynomial}
  C(s)=\sum_{i=0}^{7}a_i\,\bar{s}^i\,,
\end{align}
where $\bar{s}$ denotes the dimensionless value of $s$ measured in units of $\mathrm{GeV}^2$, and $a_i$ are the fitted coefficients given by TABLE~II of the supplementary material of Ref.~\cite{Bischofberger:2011pw}. Here we will adopt directly Eq.~\eqref{eq:polynomial} in our numerical analysis.\footnote{We thank Hisaki Hayashii for useful communications about the parametrization of the function $C(s)$.} With other different conventions taken into account, our final expression of the $CP$-violating observable $A_i^{CP}$ that connects the theoretical prediction with the Belle measurement is then given by
\begin{align}\label{eq:finalACP}
 A_i^{CP}\simeq&\Delta_{K\pi}\,S_\mathrm{EW}\,\frac{N_s}{n_i}\int_{s_{1,i}}^{s_{2,i}}\Bigg\{-\frac{\mathrm{Im}[\hat{\epsilon}_S]}{m_\tau(m_s-m_u)}\,\mathrm{Im}\left[F_+(s)F_0^\ast(s)\right] - \frac{2\mathrm{Im}[\hat{\epsilon}_T]}{m_\tau}\,\mathrm{Im}\left[F_T(s)F_0^\ast(s)\right]\nonumber\\[0.2cm]
 &\hspace{-0.7cm} +\left[\left(\frac{1}{s}+\frac{\mathrm{Re}[\hat{\epsilon}_S]}{m_\tau(m_s-m_u)}\right)\,\mathrm{Re}\left[F_+(s)F_0^\ast(s)\right]-\frac{2\mathrm{Re}[\hat{\epsilon}_T]}{m_\tau}\, \mathrm{Re}[F_T(s)F_0^\ast(s)]\right]A^{CP}_K\Bigg\}\,C(s)\,ds\,.
\end{align}
Note that the SM predictions for the $CP$-violating angular asymmetries $A_{\mathrm{SM},i}^{CP}$ given in Ref.~\cite{Chen:2020uxi} have been made by neither taking into account the detector efficiencies of the Belle experiment~\cite{Bischofberger:2011pw} nor being expressed in terms of the angle $\theta$, and should be therefore updated by using Eq.~\eqref{eq:finalACP} with all the non-standard NP couplings set to zero. The resulting numerical results are given in the second column of Table~\ref{tab:belle}. For convenience and as a comparison, the Belle measurements of the $CP$ asymmetries $A_{{\rm exp},i}^{CP}$ as well as $n_i/N_s$ in four different $K\pi$ invariant-mass bins are also listed in the third and the fourth column of Table~\ref{tab:belle}, respectively.\footnote{Here no $CP$ asymmetry is assumed in the background, and both the measured $CP$ asymmetries and the observed number of signal events in each mass bin correspond to the ones after subtracting the background contributions~\cite{Bischofberger:2011pw}.} Obviously, our central values of the SM predictions in each mass bin are even smaller than the ones obtained in Ref.~\cite{Chen:2020uxi}. Nonetheless, the main conclusion of this work is still consistent with that made in Ref.~\cite{Chen:2020uxi}, since the SM predictions obtained in these two cases are still below the current Belle detection sensitivity of $\mathcal{O}(10^{-3})$. Our SM predictions are, however, expected to be detectable at the Belle II experiment, where $\sqrt{70}$ times more sensitive results will be obtained with a $50~\mathrm{ab}^{-1}$ data sample~\cite{Kou:2018nap}.

\begin{table}[t]
	\tabcolsep 0.285in
	\renewcommand\arraystretch{1.55}
	\begin{center}
		\caption{\small Updated SM predictions including the detector efficiencies (the second column) and the Belle measurements (the third column) of the $CP$ asymmetries $A_{i}^{CP}$, as well as the observed number of signal events $n_i$ per mass bin divided by the number of total events $N_s$ (the fourth column) for four different $K\pi$ invariant-mass bins (the first column)~\cite{Bischofberger:2011pw}. \label{tab:belle} }
		\vspace{0.2cm}
		\begin{tabular}{cccc}
			\hline\hline
			$\sqrt{s}$~[GeV] & $A_{\mathrm{SM},i}^{CP}~[10^{-3}]$ & $A_{\mathrm{exp},i}^{CP}~[10^{-3}]$ & $n_i/N_s~[\%]$\\
			\hline
			$0.625-0.890$ & $0.39\pm0.01$ & $\phantom{-}7.9\pm3.0\pm2.8$ & $36.53\pm0.14$\\
			$0.890-1.110$ & $0.04\pm0.01$ &$\phantom{-}1.8\pm2.1\pm1.4$ & $57.85\pm0.15$\\
			$1.110-1.420$ & $0.12\pm0.02$ & $-4.6\pm7.2\pm1.7$ & $\phantom{0}4.87\pm0.04$\\
			$1.420-1.775$ & $0.27\pm0.05$ & $-2.3\pm19.1\pm5.5$ & $\phantom{0}0.75\pm0.02$\\
			\hline\hline
		\end{tabular}
	\end{center}
\end{table}

\subsection{Constraints on the NP parameters}
\label{sec:modelindepend}

Assuming that the NP parameters to be fitted in this work (\textit{i.e.}, the imaginary parts of the non-standard scalar and tensor couplings $\mathrm{Im}[\hat{\epsilon}_S]$ and $\mathrm{Im}[\hat{\epsilon}_T]$) obey a normal distribution and following the general procedure of the least squares method, we can obtain the best-fit values of these two parameters by minimizing the $\chi^2$ function constructed in terms of the experimental measurements:
\begin{align}\label{eq:chisq}
\chi^2=\sum_{i=1}^4\left(\frac{A_{\mathrm{exp},i}^{CP}-A_{\mathrm{th},i}^{CP}}{\sigma_i}\right)^2+\left(\frac{\mathcal{B}_\mathrm{exp}^{\tau^-}-\mathcal{B}_\mathrm{th}^{\tau^-}}{\sigma_{\mathcal{B}}}\right)^2\,,
\end{align}
where $A_{\mathrm{exp},i}^{CP}$ and $A_{\mathrm{th},i}^{CP}$ stand respectively for the experimental measurement and the theoretical prediction of the $CP$ asymmetry in the $i$-th $K\pi$ invariant-mass bin, with $\sigma_i$ being the corresponding total uncertainty obtained by adding the experimental and theoretical ones in quadrature; similarly, $\mathcal{B}_\mathrm{exp}^{\tau^-}$ and $\mathcal{B}_\mathrm{th}^{\tau^-}$ denote respectively the experimental measurement and the theoretical prediction of the branching ratio of $\tau^-\to K_S\pi^-\nu_\tau$ decay, with $\sigma_{\mathcal{B}}$ being the corresponding total uncertainty. Here, for simplicity, we have neglected possible experimental correlations among the observables $A_{\mathrm{exp},i}^{CP}$ and $\mathcal{B}_\mathrm{exp}^{\tau^-}$. The numerical inputs of $A_{\mathrm{exp},i}^{CP}$ have been given already in the third column of Table~\ref{tab:belle}, while $A_{\mathrm{th},i}^{CP}$ can be obtained from Eq.~\eqref{eq:finalACP}. The branching ratio $\mathcal{B}_\mathrm{exp}^{\tau^-}=(4.04\pm0.02\pm0.13)\times10^{-3}$ is taken from Ref.~\cite{Epifanov:2007rf}, while $\mathcal{B}_\mathrm{th}^{\tau^-}$ can be calculated by integrating the differential decay width (cf. Eq.~\eqref{eq:dG2ds}) over $s$,
\begin{align}
\mathcal{B}_\mathrm{th}^{\tau^-}=\frac{1}{2\Gamma_\tau}\int_{s_{K\pi}}^{m_\tau^2}\,\frac{d\Gamma(\tau^-\to \bar{K}^0\pi^-\nu_\tau)}{ds}\,ds\,,
\end{align}
where $\Gamma_\tau=1/\tau_\tau$ is the total decay width of the $\tau$ lepton, with $\tau_\tau=290.3\times10^{-15}~\mathrm{s}$~\cite{Zyla:2020zbs}. 

\begin{figure}[t]
	\centering
	\includegraphics[width=0.55\textwidth]{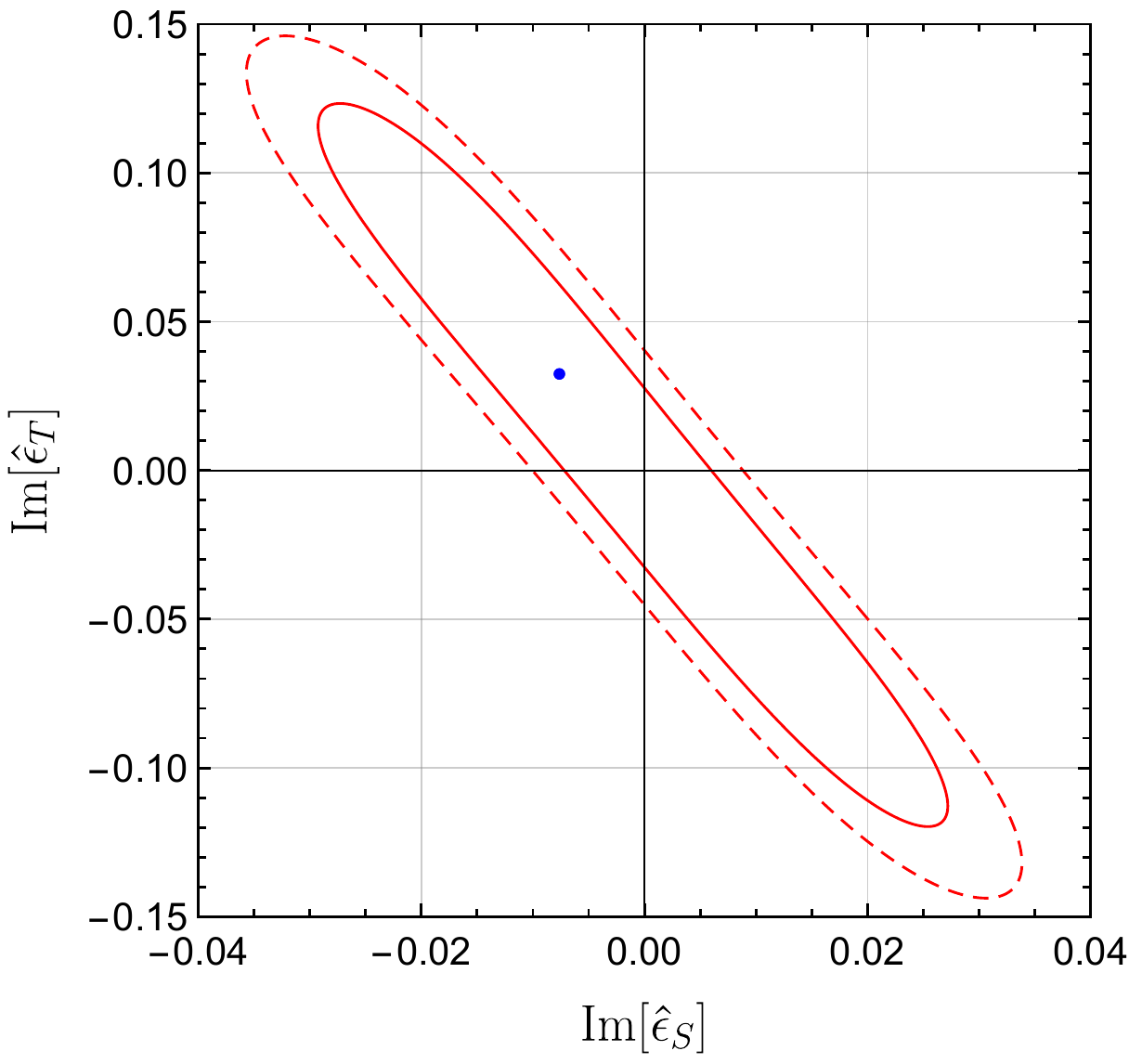}
	\caption{\small Allowed regions for the NP parameters $\mathrm{Im}[\hat{\epsilon}_S]$ and $\mathrm{Im}[\hat{\epsilon}_T]$ at the $68\%$ (region encircled by the red solid curve) and the $90\%$ C.L. (region encircled by the red dashed curve), respectively. The blue point $(-0.008,0.03)$ corresponds to the best-fit values of $\mathrm{Im}[\hat{\epsilon}_S]$ and $\mathrm{Im}[\hat{\epsilon}_T]$. \label{fig:region} }
\end{figure}

With all the numerical inputs and the theoretical expressions at hand, we can now obtain the final resulting bounds on the NP parameters $\mathrm{Im}[\hat{\epsilon}_S]$ and $\mathrm{Im}[\hat{\epsilon}_T]$, which are given, respectively, by
\begin{align}\label{eq:results}
\mathrm{Im}[\hat{\epsilon}_S]=-0.008\pm0.027,\qquad \mathrm{Im}[\hat{\epsilon}_T]=0.03\pm0.12\,,
\end{align}
with the best-fit values corresponding to a minimum $\chi^2_\mathrm{min}=4.20$ for two degrees of freedom, and the associated correlation matrix is given by
\begin{align}\label{eq:correlation}
\rho=\begin{pmatrix} 1 & -0.986 \\[0.15cm] -0.986 & 1 \end{pmatrix}\,.
\end{align}
Comparing Eq.~\eqref{eq:results} with the numerical results obtained in Refs.~\cite{Bischofberger:2011pw,Cirigliano:2017tqn}, one can see that our bound on $\mathrm{Im}[\hat{\epsilon}_S]$ is consistent with that obtained by Belle~\cite{Bischofberger:2011pw}, $|\mathrm{Im}(\eta_S)|<0.026$, at $90\%$ C.L., while the upper limit on $\mathrm{Im}[\hat{\epsilon}_T]$ is only of $\mathcal{O}(10^{-1})$, being therefore not competitive with the bound $2|\mathrm{Im}[\hat{\epsilon}_T]|\lesssim 10^{-5}$ obtained from the neutron EDM and $D^0-\bar{D}^0$ mixing~\cite{Cirigliano:2017tqn}, which is attributed to the large uncertainties of the current Belle measurements of the $CP$ asymmetries $A_{\mathrm{exp},i}^{CP}$~\cite{Bischofberger:2011pw}. However, the numerical bounds given by Eq.~\eqref{eq:results} are expected to be improved with more precise measurements from the Belle II experiment~\cite{Kou:2018nap}, as well as the proposed Tera-Z~\cite{Pich:2020qna} and STCF~\cite{Sang:2020ksa} facilities. To show further the correlation between the two NP parameters $\mathrm{Im}[\hat{\epsilon}_S]$ and $\mathrm{Im}[\hat{\epsilon}_T]$, we plot in Figure~\ref{fig:region} their allowed regions at the $68\%$ (region encircled by the red solid curve) and the $90\%$ C.L. (region encircled by the red dashed curve), respectively. It can be seen from either the correlation matrix given by Eq.~\eqref{eq:correlation} or Figure~\ref{fig:region} that there exists a remarkably negative correlation between $\mathrm{Im}[\hat{\epsilon}_S]$ and $\mathrm{Im}[\hat{\epsilon}_T]$. This is attributed to the fact that the $K\pi$ vector and tensor form factors are both dominated by the vector resonances $K^\ast(892)$ and $K^\ast(1410)$, and hence have almost the same phases, especially in the elastic  region~\cite{Cirigliano:2017tqn}, making their contributions to the $CP$ asymmetries different from each other only by a different normalization factor, as demonstrated explicitly by Eq.~\eqref{eq:finalACP}. Although the form-factor phases $\delta_T(s)$ and $\delta_+(s)$ start to behave differently in the inelastic region due to the different relative weights of the two resonances (cf. Figure~\ref{fig:FFs}), such an effect will be diluted by the larger experimental uncertainties of the $CP$ asymmetries in the higher bins, as can be seen from the third column of Table~\ref{tab:belle}. Thus, the negative correlation between $\mathrm{Im}[\hat{\epsilon}_S]$ and $\mathrm{Im}[\hat{\epsilon}_T]$ always remains in the four $K\pi$ invariant-mass bins, even with the uncertainty of the tensor form-factor phase (relative to the vector one) in the inelastic region taken into account.

\begin{figure}[t]
	\centering
	\includegraphics[width=0.48\textwidth]{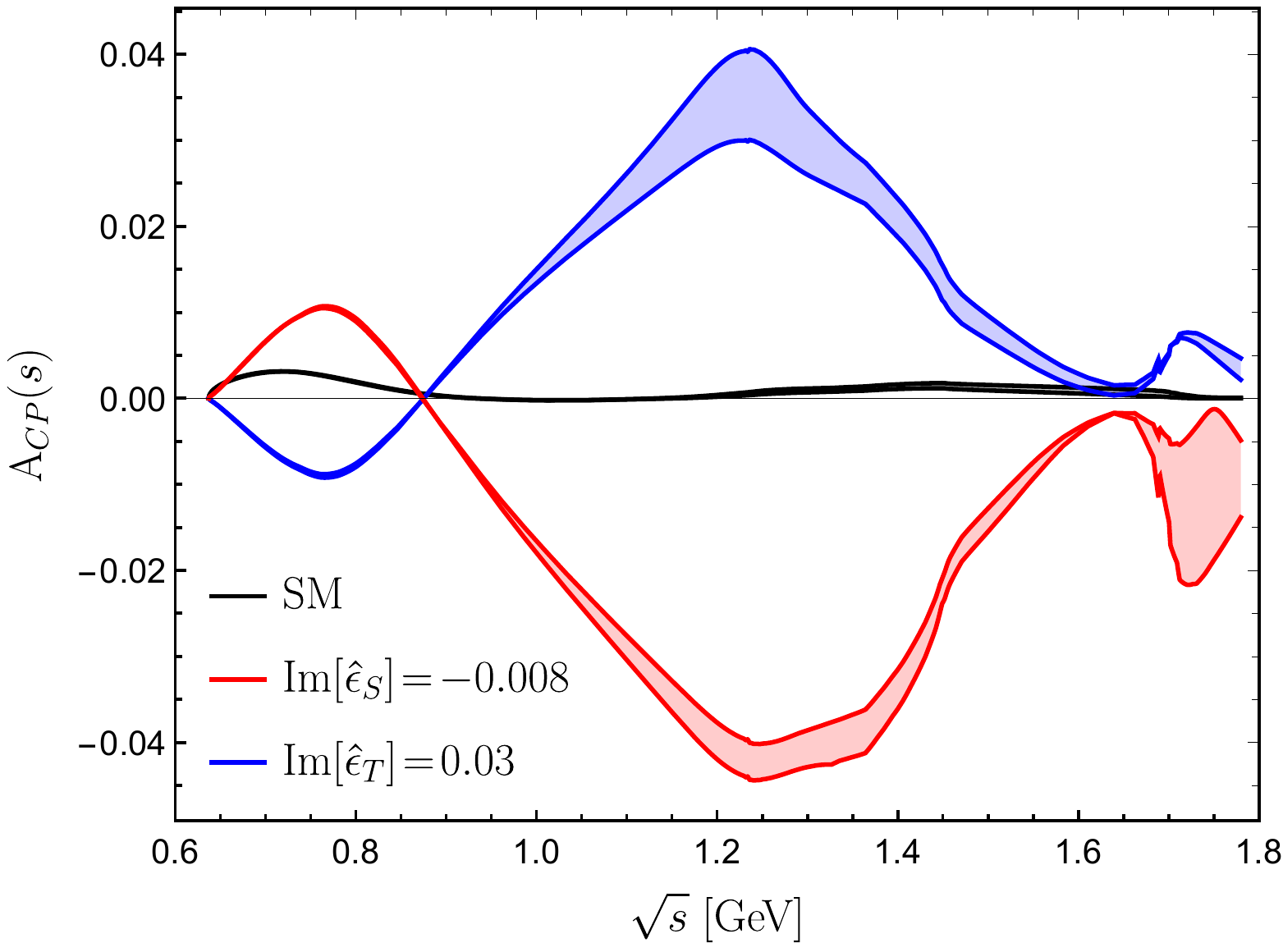}
	\hspace{0.12in}
	\includegraphics[width=0.48\textwidth]{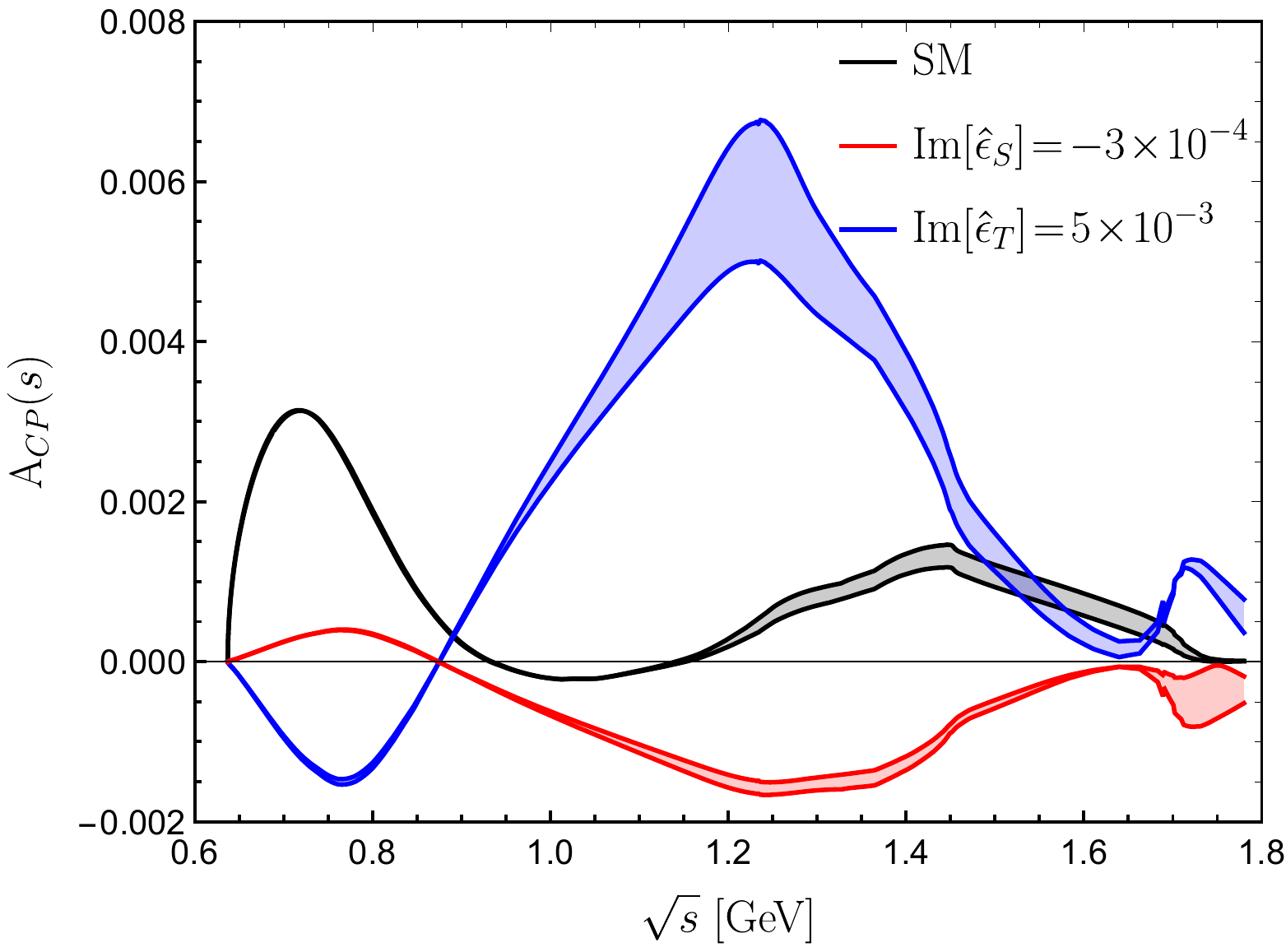}
	\caption{\small Left: Distributions of the $CP$ asymmetries in the whole $K\pi$ invariant-mass region for three different cases, the SM prediction (gray band), the non-standard scalar contribution with $\mathrm{Im}[\hat{\epsilon}_S]=-0.008$ (red band), and the non-standard tensor contribution with $\mathrm{Im}[\hat{\epsilon}_T]=0.03$ (blue band). Right: The zoomed-in version of the SM prediction as well as the cases with $\mathrm{Im}[\hat{\epsilon}_S]=-3\times10^{-4}$ (red band) and $\mathrm{Im}[\hat{\epsilon}_T]=5\times10^{-3}$ (blue band). \label{fig:ACP} }
\end{figure}

Finally, in order to compare the NP contributions with the SM expectation for the $CP$ asymmetry in the angular distributions of $\tau\to K_S\pi\nu_\tau$ decays, we also plot in Figure~\ref{fig:ACP} the distributions of the $CP$ asymmetries in the whole $K\pi$ invariant-mass region, with three different cases: the SM prediction induced by the indirect CPV in $K^0-\bar{K}^0$ mixing (gray band)~\cite{Chen:2020uxi}, the non-standard scalar contribution with the best-fit value $\mathrm{Im}[\hat{\epsilon}_S]=-0.008$ (red band), and the non-standard tensor contribution with the best-fit value $\mathrm{Im}[\hat{\epsilon}_T]=0.03$ (blue band). Here we have taken into account both the theoretical (including especially the systematic uncertainty of the tensor form factor induced by varying the parameter $\beta$ between $\beta=+0.75\gamma$ and $\beta=-0.75\gamma$; see the appendix for further details) and the experimental uncertainties of the input parameters. It can be seen that, when using the best-fit values of $\mathrm{Im}[\hat{\epsilon}_S]$ and $\mathrm{Im}[\hat{\epsilon}_T]$ as inputs, the distributions of the $CP$ asymmetries have almost the same magnitude but are opposite in sign in the whole $K\pi$ invariant-mass region, and the maximum absolute values reached at around $\sqrt{s}=1.2~\mathrm{GeV}$ for both the non-standard scalar and tensor contributions are about one order of magnitude larger than that of the SM prediction. This implies that the $CP$ asymmetry in the angular distributions of $\tau\to K_S\pi\nu_\tau$ decays could be significantly enhanced if these kinds of NP contributions are present. As a consequence, we strongly recommend more precise studies of the $CP$ asymmetry in the angular distributions of $\tau\to K_S\pi\nu_\tau$ decays from both the theoretical and experimental aspects, to further test the observations made in this work.

\subsection{Bounds on the NP parameters from other processes}
\label{subsec:otherbounds}

If the non-standard scalar and tensor interactions contributing to the $\tau\to K_S\pi\nu_\tau$ decays are assumed to originate from a weakly-coupled heavy NP well above the electroweak scale, the $SU(2)_L$ invariance of the resulting SMEFT Lagrangian~\cite{Buchmuller:1985jz,Grzadkowski:2010es,Brivio:2017vri} would indicate that very strong limits on the imaginary parts of the non-standard scalar and tensor coefficients, $\mathrm{Im}[\hat{\epsilon}_S]$ and $\mathrm{Im}[\hat{\epsilon}_T]$, could also be obtained from the neutron EDM and the $D^0-\bar{D}^0$ mixing~\cite{Cirigliano:2017tqn}. In this subsection, we will explore the maximum effects on the $CP$ asymmetries in $\tau\to K_S\pi\nu_\tau$ decays, by taking into account the bounds from the neutron EDM and the $D^0-\bar{D}^0$ mixing.

\subsubsection{Phenomenological constraints on $\mathrm{Im}[\hat{\epsilon}_T]$}

Let us firstly discuss the phenomenological constraints on the imaginary part of the non-standard tensor coefficient $\mathrm{Im}[\hat{\epsilon}_T]$. As demonstrated already in Ref.~\cite{Cirigliano:2017tqn}, the coefficient $\mathrm{Im}[\hat{\epsilon}_T]$, which can provide a non-vanishing contribution to the $CP$ asymmetries of $\tau\to K_S\pi\nu_\tau$ decays (cf. Eqs.~\eqref{eq:intACP} and \eqref{eq:finalACP}), should also be subject to the stringent constraints from the neutron EDM and the $D^0-\bar{D}^0$ mixing, because above the electroweak scale the tensor operator $(\bar{\nu}_\tau\sigma_{\mu\nu}\tau_R)(\bar{s}_L\sigma^{\mu\nu}u_R)$ relevant for the $\tau\to K_S\pi\nu_\tau$ decays originates from the following $SU(3)_C\times SU(2)_L\times U(1)_Y$ gauge-invariant SMEFT Lagrangian~\cite{Buchmuller:1985jz,Grzadkowski:2010es,Brivio:2017vri}:
\begin{align}\label{eq:SMEFTLT}
\mathcal{L}_\mathrm{SMEFT} &\supset [C^{(3)}_{\ell equ}]_{klmn} (\bar{\ell}^i_{Lk}\sigma_{\mu\nu}e_{Rl})\epsilon^{ij}(\bar{q}_{Lm}^{j}\sigma^{\mu\nu}u_{Rn})+{\rm h.c.}\, \nonumber \\[0.2cm]
& \hspace{-0.6cm} =[C^{(3)}_{\ell equ}]_{klmn}\big[(\bar{\nu}_{L k} \sigma_{\mu \nu} e_{R l})(\bar{d}_{L m} \sigma^{\mu \nu} u_{R n}) - (\bar{e}_{L k} \sigma_{\mu \nu} e_{R l})(\bar{u}_{L m} \sigma^{\mu \nu} u_{R n})\big]+\text{h.c.}\,,
\end{align}
where $\ell_L=(\nu_L,e_L)^{T}$ and $q_L=(u_L,d_L)^{T}$ denote the left-handed lepton and quark $SU(2)_L$ doublets, while $e_R$ and $u_R$ are the right-handed charged lepton and up-quark $SU(2)_L$ singlets, with $i$, $j$ being the $SU(2)_L$ indices and $k$, $l$, $m$, $n$ the generation indices. Transforming from the gauge to the mass basis for fermions, we can rewrite Eq.~\eqref{eq:SMEFTLT} as
\begin{equation}\label{eq:LTmassbasis}
\mathcal{L}_\mathrm{SMEFT} \supset [C^{(3)}_{\ell equ}]_{klmn} \big[(\bar{\nu}_{Lk}\sigma_{\mu\nu}e_{Rl})(\bar{d}_{Lm}\sigma^{\mu\nu}u_{Rn})-V_{am}(\bar{e}_{Lk}\sigma_{\mu\nu}e_{Rl})(\bar{u}_{La}\sigma^{\mu\nu}u_{Rn})\big]+{\rm h.c.}\,,
\end{equation}
where, without loss of generality, we have chosen the \textit{down basis}, in which both the down-quark and the charged lepton Yukawa couplings are diagonal, while the right-handed fermions are in the mass basis, with $V_{am}$ being the CKM quark-mixing matrix. Note that the $C$ coefficients in Eq.~\eqref{eq:LTmassbasis}, which are now given in the fermion mass basis, are obtained as bi-unitary transformations of the corresponding ones in Eq.~\eqref{eq:SMEFTLT} defined in the gauge basis. The non-standard tensor coefficient $\hat{\epsilon}_T$ defined in Eq.~\eqref{eq:Efective_Lagrangian} is then related to the $C$ coefficient in Eq.~\eqref{eq:LTmassbasis} via the tree-level relation at the electroweak scale
\begin{align}\label{eq:CandT}
	[C_{\ell e q u}^{(3)}]_{3321}=-2 \sqrt{2} G_{F} V_{u s} \hat{\epsilon}_{T}^{*}\,.
\end{align}

\begin{figure}[t]
	\centering
	\includegraphics[width=0.45\textwidth]{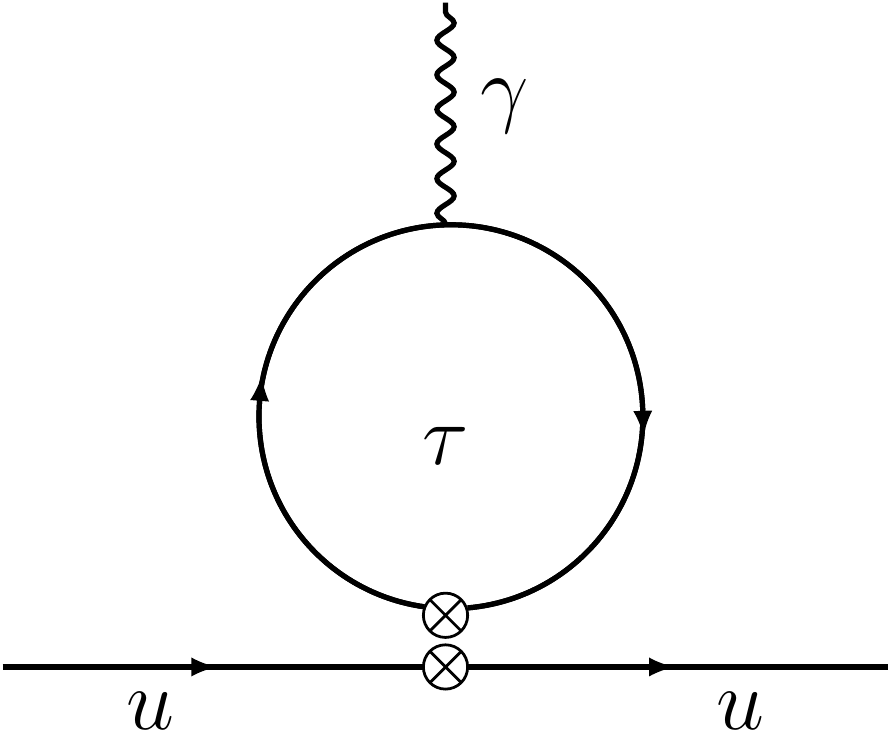}
	\hspace{0.3in}
	\includegraphics[width=0.45\textwidth]{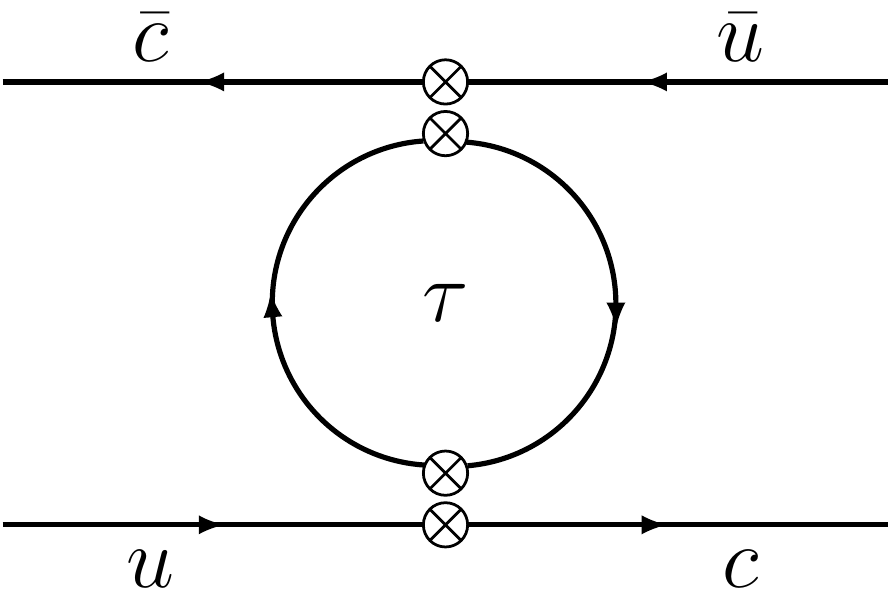}
	\caption{\small Diagrammatic representations of the electromagnetic dipole operator contributing to the neutron EDM induced by inserting the operator $(\bar{\tau}_L\sigma_{\mu\nu}\tau_R)(\bar{u}_L\sigma^{\mu\nu}u_R)$ (left), and the $\Delta C=2$ four-quark operators contributing to the $D^0-\bar{D}^0$ mixing produced by a double insertion of the operator $(\bar{\tau}_L\sigma_{\mu\nu}\tau_R)(\bar{c}_L\sigma^{\mu\nu}u_R)$ (right, and its permutation should also be taken into account). \label{fig:nEDM&Dmixing} }
\end{figure}

It is clear from Eq.~\eqref{eq:LTmassbasis} that the operator $(\bar{\tau}_L\sigma_{\mu\nu}\tau_R)(\bar{u}_L\sigma^{\mu\nu}u_R)$, which has a contribution to the neutron EDM through the renormalization group (RG) evolution~\cite{Jenkins:2013zja,Jenkins:2013wua,Jenkins:2017jig,Jenkins:2017dyc,Cirigliano:2017azj,Crivellin:2019qnh}, shares the same Wilson coefficient $[C^{(3)}_{\ell equ}]_{3321}$ with the tensor operator $(\bar{\nu}_\tau\sigma_{\mu\nu}\tau_R)(\bar{s}_L\sigma^{\mu\nu}u_R)$ that contributes to the $\tau\to K_S\pi\nu_\tau$ decays, up to an additional CKM matrix element $V_{us}$. This implies that the tensor coefficient $\mathrm{Im}[\hat{\epsilon}_T]$ receives also the constraint from the neutron EDM~\cite{Cirigliano:2017tqn}. Explicitly, the operator $(\bar{\tau}_L\sigma_{\mu\nu}\tau_R)(\bar{u}_L\sigma^{\mu\nu}u_R)$ contributes to the up-quark EDM $d_u(\mu)$ via the left diagram shown in Figure~\ref{fig:nEDM&Dmixing}, with the result given by~\cite{Cirigliano:2017tqn,Bobeth:2011st}
\begin{align}
\mathcal{L}_D=-\frac{i}{2}d_u(\mu)\bar{u}\sigma^{\mu\nu}\gamma_5u F_{\mu\nu}\,,
\end{align} 
which is related to the neutron EDM via 
\begin{align}
d_n=g^u_T(\mu) d_u(\mu)\,,
\end{align}
where $g^u_T(2\,\text{GeV})=-0.204(11)(10)$~\cite{Gupta:2018lvp,Bhattacharya:2015esa} is the flavour-diagonal tensor charge of the neutron given in the $\mathrm{\overline{MS}}$ scheme. Making use of the current $90\%$ C.L. bound on the neutron EDM, $|d_n|<1.8\times10^{-26}\,e\,{\rm cm}$~\cite{nEDM:2020crw}, one can then obtain a stringent limit on $\mathrm{Im}[\hat{\epsilon}_T]$~\cite{Cirigliano:2017tqn}.  

It should be noted, however, that the coefficient associated with the tensor operator $(\bar{\tau}_L\sigma_{\mu\nu}\tau_R)(\bar{u}_L\sigma^{\mu\nu}u_R)$ could originate either from the single $V_{us}\,[C_{\ell e q u}^{(3)}]_{3321}$ or the combination $V_{ud}\,[C_{\ell e q u}^{(3)}]_{3311}+V_{us}\,[C_{\ell e q u}^{(3)}]_{3321}$ (cf. Eq.~\eqref{eq:LTmassbasis}).\footnote{Here the term proportional to the much smaller CKM matrix element $V_{ub}$ has been neglected safely. It should also be pointed out that the combination, $V_{us}\left(V_{ud}\,\mathrm{Im}[c_T^{11}]+V_{us}\,\mathrm{Im} [c_T^{21}]\right)$, has been misused in Ref.~\cite{Cirigliano:2017tqn} and it should be replaced by the correct one, $V_{ud}^2\,\mathrm{Im}[c_T^{11}]+V_{us}^2\,\mathrm{Im} [c_T^{21}]$; this can lead to a slightly different plot as depicted in Figure~\ref{fig:com}.} Thus, the stringent bound on $\mathrm{Im}[\hat{\epsilon}_T]$ obtained in Ref.~\cite{Cirigliano:2017tqn} could be diluted when there exists an extraordinary cancellation between the two terms in the combination 
\begin{equation}\label{eq:combine}
	V_{ud}\,\mathrm{Im}[C^{(3)}_{\ell equ}]_{3311}+V_{us}\,\mathrm{Im}[C^{(3)}_{\ell equ}]_{3321}=2\sqrt{2}G_F\left(V^2_{ud}\,\mathrm{Im}[\epsilon_T]_{3311}+V^2_{us}\,\mathrm{Im}[\epsilon_T]_{3321}\right)\,,
\end{equation}
where $\hat{\epsilon}_T$ is equal to $[\epsilon_T]_{3321}^*$ working at the linear order in the $\epsilon_i$ coefficients defined in Eq.~\eqref{eq:Efective_Lagrangian}. Let us discuss these two cases separately.
\begin{enumerate}
\item[(i)] In the case of a ``single coefficient dominance'' assumption, \textit{i.e.}, the neutron EDM receives contribution only from the single tensor coefficient $\hat{\epsilon}_T$, by making use of the $90\%$ C.L. bound on the neutron EDM, $|d_n|<1.8\times10^{-26}\,e\,{\rm cm}$~\cite{nEDM:2020crw}, as well as the solution of the RG evolution for the up-quark EDM,
\begin{align}\label{eq:du}
d_u(\mu)=-2\sqrt{2}G_F\,\frac{e m_{\tau}}{\pi^2}\,V_{us}^2\,\mathrm{Im}[\hat{\epsilon}_T(\mu)] \log\frac{\Lambda}{\mu}\,,
\end{align}
we obtain a very stringent bound, $|\mathrm{Im}[\hat{\epsilon}_T(\mu_\tau)]| \leq 1.5\times 10^{-5}/\log(\Lambda/\mu_\tau)\lesssim 4\times10^{-6}$, for the scales $\Lambda\gtrsim100~{\rm GeV}$ and $\mu_\tau=2~\text{GeV}$.\footnote{The tensor coefficient $c_T$ in Ref.~\cite{Cirigliano:2017tqn} is related to $\hat{\epsilon}_T$ in our notation via $c_T=2\,\hat{\epsilon}_T^*$. This, together with the updated inputs of $|d_n|$ and $g_T^u(2~\text{GeV})$ listed in Table~\ref{tab:input}, leads to an even stronger limit on $|\mathrm{Im}[\hat{\epsilon}_T(\mu_\tau)]|$ than obtained in Ref.~\cite{Cirigliano:2017tqn}.} This is, as far as we know, the strongest limit on $\mathrm{Im}[\hat{\epsilon}_T]$ obtained so far, and should be taken into account for any realistic estimate of the $CP$ asymmetries in $\tau\to K_S\pi\nu_\tau$ decays, once the tensor operator is assumed to originate from a weakly-coupled heavy NP well above the electroweak scale. 

With such a stringent limit on $\mathrm{Im}[\hat{\epsilon}_T]$, we can then explore the maximum effects on the $CP$ asymmetries in $\tau\to K_S\pi\nu_\tau$ decays. Substituting the value $\mathrm{Im}[\hat{\epsilon}_T]=4\times10^{-6}$ into Eq.~\eqref{eq:finalACP} and neglecting other contributions, we obtain a $CP$ asymmetry distributed in the whole $K\pi$ invariant-mass region that is of the same shape as the blue band shown in the left panel of Figure~\ref{fig:ACP}, but with the magnitude of its extreme value being only at the level of $\mathcal{O}(10^{-6})$, which is three orders of magnitude smaller than the SM prediction ($\mathcal{O}(10^{-3})$) and can be therefore neglected safely.

\item[(ii)] In the case when there exists an extraordinary cancellation between the two terms of Eq.~\eqref{eq:combine}, the up-quark EDM given by Eq.~\eqref{eq:du} will be modified as
\begin{align}\label{eq:dumodi}
d_u(\mu)=-2\sqrt{2}G_F\,\frac{e m_{\tau}}{\pi^2}\left(V^2_{ud}\,\mathrm{Im}[\epsilon_T]_{3311}+V^2_{us}\,\mathrm{Im}[\epsilon_T]_{3321}\right)\log\frac{\Lambda}{\mu}\,.
\end{align}
As a consequence, the stringent bound on $\mathrm{Im}[\hat{\epsilon}_T]$ obtained in case~(i) could be diluted. Nevertheless, one has to consider in this case another combination, $V_{cd}\,[C^{(3)}_{\ell equ}]_{3311}+V_{cs}\,[C^{(3)}_{\ell equ}]_{3321}=V_{ud}V_{cd}\,[\epsilon_T]_{3311}+V_{us}V_{cs}\,[\epsilon_T]_{3321}$, the imaginary part of which will be subject to the constraint from $D^0-\bar{D}^0$ mixing~\cite{Cirigliano:2017tqn}.\footnote{Again, the term proportional to the smaller CKM matrix element $V_{cb}$ has been neglected, and the misuse of the combination $V_{us}\left(V_{cd}\,c_T^{11}+V_{cs}\,c_T^{21}\right)$ in Ref.~\cite{Cirigliano:2017tqn} should be corrected by $V_{cd}V_{ud}\,c_T^{11}+V_{cs}V_{us}\,c_T^{21}$.} 

After a double insertion of the operator $(\bar{\tau}_L\sigma_{\mu\nu}\tau_R)(\bar{c}_L\sigma^{\mu\nu}u_R)$ into the right diagram of Figure~\ref{fig:nEDM&Dmixing} and a proper Fierz rearrangement, one arrives at the $\Delta C=2$ effective Hamiltonian describing the $D^0-\bar{D}^0$ mixing~\cite{UTfit:2007eik,Bazavov:2017weg}:
\begin{align}\label{eq:Heffo2o3}
\mathcal{H}_{\rm eff}^{\Delta C=2}=C_2^\prime(\bar{c}_L^\alpha u_R^\alpha)(\bar{c}_L^\beta u_R^\beta)+C_3^\prime(\bar{c}_L^\alpha u_R^\beta)(\bar{c}_L^\beta u_R^\alpha)\,,
\end{align}
where $\alpha$ and $\beta$ denote the colour indices, and the resulting short-distance Wilson coefficients are given by~\cite{Cirigliano:2017tqn}
\begin{align}\label{eq:C2C3}
C^\prime_2=\frac{1}{2}C^\prime_3=16G_F^2\,\frac{m_\tau^2}{\pi^2}\left(V_{ud}V_{cd}\,[\epsilon_T]_{3311}+V_{us}V_{cs}\,[\epsilon_T]_{3321}\right)^2\,\log\frac{\Lambda}{\mu_\tau}\,.
\end{align}
Here we have neglected the masses of the external charm and up quarks during the calculation. Starting with the $\Delta C=2$ effective Hamiltonian defined by Eq.~\eqref{eq:Heffo2o3}, one can obtain the off-diagonal element of the mass matrix,
\begin{equation}\label{eq:M12}
	M_{12}^{\text{NP}}=\frac{1}{2M_D}\Big[C^\prime_2(\mu)\langle D^0|(\bar{c}_L^\alpha u_R^\alpha)(\bar{c}_L^\beta u_R^\beta)|\bar{D}^0\rangle(\mu) + C^\prime_3(\mu)\langle D^0|(\bar{c}_L^\alpha u_R^\beta)(\bar{c}_L^\beta u_R^\alpha)|\bar{D}^0\rangle(\mu)\Big]\,,
\end{equation}
and the ``theoretical'' mixing parameters of the neutral $D$-meson system~\cite{Kagan:2009gb,Amhis:2019ckw}, 
\begin{align}\label{eq:x12}
	x_{12}^{\text{NP}}=\frac{2|M_{12}^{\text{NP}}|}{\Gamma_D}\,, \qquad \phi_{12}^{\text{NP}}={\rm arg}\left(\frac{M_{12}^{\text{NP}}}{\Gamma_{12}}\right)\,,
\end{align}
where $M_D$ and $\Gamma_D$ stand respectively for the averaged mass and decay width of the neutral $D$ meson, whereas the off-diagonal element of the decay matrix, $\Gamma_{12}$, will be assumed to be unaffected by the NP contribution.\footnote{Here we have assumed that the NP does not change the phase of $\Gamma_{12}$ and, as $M_{12}$ and $\Gamma_{12}$ are both real to a very good approximation within the SM, the relative phase $\phi_{12}^{\text{NP}}$ can be treated as the phase of $M_{12}^{\text{NP}}$.} The short-distance hadronic matrix elements of the $\Delta C=2$ four-quark operators in Eq.~\eqref{eq:M12} have been evaluated, \textit{e.g.}, by the FNAL/MILC collaboration~\cite{Bazavov:2017weg}. Confronting the NP contributions given by Eqs.~\eqref{eq:M12}--\eqref{eq:x12} with the latest global fit for the charm-mixing data performed by the Heavy Flavor Averaging Group (HFLAV)~\cite{Amhis:2019ckw}, 
\begin{align} \label{eq:expx12phi12}
	x_{12}=(0.409\pm0.048)\%\,,\qquad \phi_{12}=\left(0.58^{+0.91}_{-0.90}\right)^\circ\,,
\end{align}
one can then obtain constraints on the tensor coefficients $[\epsilon_T]_{3311}$ and $[\epsilon_T]_{3321}$~\cite{Cirigliano:2017tqn}.  

\begin{figure}[t]
	\centering
	\includegraphics[width=0.98\textwidth]{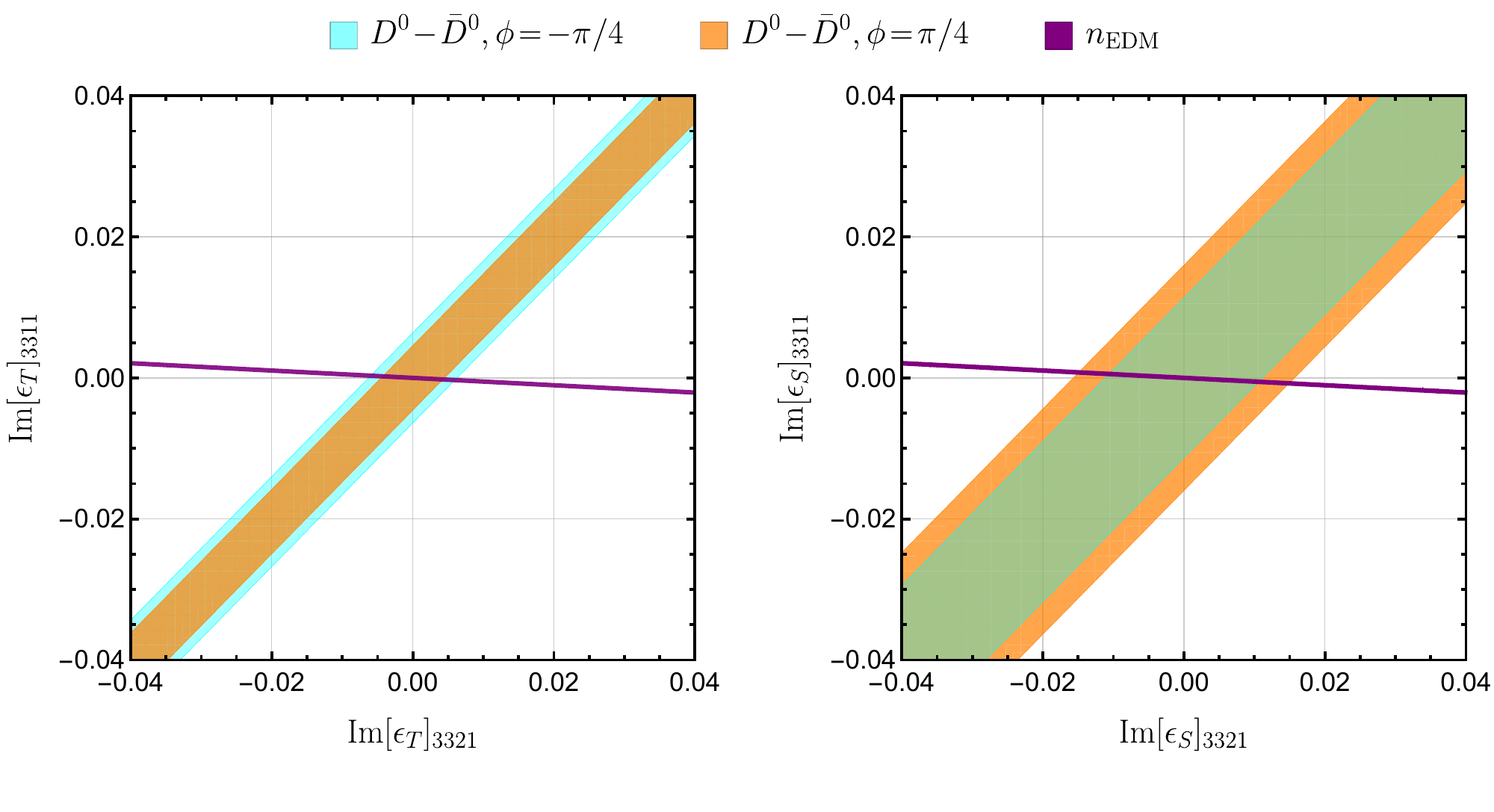}
	\vspace{-0.4cm}
	\caption{\small Allowed regions for the imaginary parts of the tensor (left) and scalar (right) coefficients from the neutron EDM (purple) and the $D^0-\bar{D}^0$ mixing with $\phi=-\pi/4$ (cyan) and $\phi=\pi/4$ (orange), where the NP scale is taken at $\Lambda=1~\text{TeV}$. Here $\phi$ is the phase of $V_{ud}V_{cd}\,[\epsilon_{T(S)}]_{3311}+V_{us}V_{cs}\,[\epsilon_{T(S)}]_{3321}$, and the choices $\phi=\pm\pi/4$ are to ensure that the NP Wilson coefficients $C^\prime_{2,3}$ are purely imaginary~\cite{Cirigliano:2017tqn}. \label{fig:com} }
\end{figure}

Combining Eqs.~\eqref{eq:dumodi}, \eqref{eq:M12} and \eqref{eq:x12} with the $90\%$ C.L. bound on the neutron EDM, $|d_n|<1.8\times10^{-26}\,e\,{\rm cm}$~\cite{nEDM:2020crw}, and the fitted results for the $D^0-\bar{D}^0$ mixing parameters given by Eq.~\eqref{eq:expx12phi12}, one can obtain straightforwardly the allowed regions for the tensor coefficients $\mathrm{Im}[\epsilon_T]_{3311}$ and $\mathrm{Im}[\epsilon_T]_{3321}$. To maximize the constraint from $D^0-\bar{D}^0$ mixing, we assume that the NP Wilson coefficients $C^\prime_{2,3}$ in Eq.~\eqref{eq:C2C3} are purely imaginary, which implies that the phase of $V_{ud}V_{cd}\,[\epsilon_T]_{3311}+V_{us}V_{cs}\,[\epsilon_T]_{3321}$ is equal to $\phi=\pm\pi/4$~\cite{Cirigliano:2017tqn}.\footnote{For the most general case, the constraint is diluted by
$\sqrt{|\tan\phi|}$ and thus disappears for $\phi=\pm\pi/2$~\cite{Cirigliano:2017tqn}.} This leads to the plot shown in the left panel of Figure~\ref{fig:com}. It should be stressed that the plot is just an updated version of FIG.~4 in Ref.~\cite{Cirigliano:2017tqn}, obtained after correcting the combinations $V_{us}\left(V_{ud}\,c_T^{11}+V_{us}\,c_T^{21}\right)$ with $V_{ud}^2\,c_T^{11}+V_{us}^2\,c_T^{21}$ (for the neutron EDM) and $V_{us}\left(V_{cd}\,c_T^{11}+V_{cs}\,c_T^{21}\right)$ with $V_{cd}V_{ud}\,c_T^{11}+V_{cs}V_{us}\,c_T^{21}$ (for the $D^0-\bar{D}^0$ mixing), and using the updated data on the neutron EDM~\cite{nEDM:2020crw} and the $D^0-\bar{D}^0$ mixing~\cite{Amhis:2019ckw}. One can see that the upper limit on $|\mathrm{Im}[\hat{\epsilon}_T]|$ in this case is now restricted to be about $5\times10^{-3}$, being therefore significantly diluted relative to that obtained in case (i). Such a diluted bound is, however, still two orders of magnitude smaller than the value $\mathrm{Im}[\hat{\epsilon}_T]\simeq\mathcal{O}(10^{-1})$ required to explain the $2.8\sigma$ discrepancy between the SM prediction~\cite{Bigi:2005ts,Calderon:2007rg,Grossman:2011zk,Chen:2019vbr} and the BaBar measurement~\cite{BABAR:2011aa} of the decay-rate asymmetry. Nevertheless, even with $|\mathrm{Im}[\hat{\epsilon}_T]|\simeq5\times10^{-3}$, there still exists observable effect on the $CP$-violating angular observable, due to its higher sensitivity to the non-standard tensor contribution than does the decay-rate asymmetry. To this end, let us make a simple estimate by choosing $\mathrm{Im}[\hat{\epsilon}_T]=5\times10^{-3}$, and the resulting $A_{CP}(s)$ is shown by the blue band in the right plot of Figure~\ref{fig:ACP}. It can be seen that, in the case when there exists an extraordinary cancellation between the two terms of Eq.~\eqref{eq:combine} with a diluted $|\mathrm{Im}[\hat{\epsilon}_T]|$, the non-standard tensor interaction still has a significant effect on the $CP$ asymmetry in the angular distributions of $\tau\to K_S\pi\nu_\tau$ decays, being especially larger than the SM prediction at around $\sqrt{s}=1.2~\mathrm{GeV}$.
\end{enumerate}

\subsubsection{Phenomenological constraints on $\mathrm{Im}[\hat{\epsilon}_S]$}

Following the same procedure as for $\mathrm{Im}[\hat{\epsilon}_T]$, we can obtain the phenomenological constraints on the imaginary part of the scalar coefficient $\mathrm{Im}[\hat{\epsilon}_S]$. The low-energy scalar operator in Eq.~\eqref{eq:Efective_Lagrangian} can originate from the following two SMEFT operators~\cite{Buchmuller:1985jz,Grzadkowski:2010es,Brivio:2017vri}: 
\begin{align}\label{eq:SMEFTLS}
\mathcal{L}_\mathrm{SMEFT} \supset [C^{(1)}_{\ell equ}]_{klmn}(\bar{\ell}^i_{Lk}e_{Rl})\epsilon^{ij}(\bar{q}_{Lm}^{j}u_{Rn})+[C_{\ell edq}]_{klmn}(\bar{\ell}^i_{Lk}e_{Rl})(\bar{d}_{Rm}q_{Ln}^{i})+{\rm h.c.}\,,
\end{align}
which can be again rewritten in the fermion mass basis as
\begin{align}\label{eq:LSmassbasis}
\mathcal{L}_\mathrm{SMEFT} &\supset [C^{(1)}_{\ell equ}]_{klmn}\left[(\bar{\nu}_{Lk} e_{Rl})(\bar{d}_{Lm}u_{Rn})-V_{am}(\bar{e}_{Lk}e_{Rl})(\bar{u}_{La}u_{Rn})\right] \nonumber\\[0.2cm]
&+[C_{\ell edq}]_{klmn}\left[V^*_{an}(\bar{\nu}_{Lk} e_{Rl})(\bar{d}_{Rm}u_{La})+(\bar{e}_{Lk}e_{Rl})(\bar{d}_{Rm}d_{Ln})\right] +{\rm h.c.}\,,
\end{align}
where the first line produces the operator $(\bar{\nu}_\tau\tau_R)(\bar{s}_Lu_R)$, whereas the second line the operator $(\bar{\nu}_\tau\tau_R)(\bar{s}_Ru_L)$, and the sum of these two operators gives the scalar operator $(\bar{\nu}_\tau\tau_R)(\bar{s}u)$ relevant for the $\tau\to K_S\pi\nu_\tau$ decays, with their coefficients related by $-2\sqrt{2}G_FV_{us}^{\ast}\hat{\epsilon}_S^{\ast}=[C^{(1)}_{\ell equ}]_{3321}+V^*_{ud}\,[C_{\ell edq}]_{3321}+V^*_{us}\,[C_{\ell edq}]_{3322}+V^*_{ub}\,[C_{\ell edq}]_{3323}$. Thus, once the scalar operator is assumed to originate from a weakly-coupled heavy NP well above the electroweak scale, very stringent limits on the scalar coefficient $\hat{\epsilon}_S$ could also be obtained from other processes. Furthermore, the allowed values of $\hat{\epsilon}_S$ obtained could also be diluted when there exist potential cancellations between $C^{(1)}_{\ell equ}$ and $C_{\ell edq}$. Here, for simplicity, we will restrict ourselves to elaborate the case where only $C^{(1)}_{\ell equ}$ is present, while the case with $C_{\ell edq}$ alone is similar and will not be discussed any further. 

A key point that should be kept in mind here is that the scalar and tensor operators can mix into each other when performing the RG evolutions both below and above the electroweak scale~\cite{Gonzalez-Alonso:2017iyc}. This implies that, even if the scalar operator $(\bar{\nu}_\tau\tau_R)(\bar{s}_Lu_R)$ has no direct contribution to the neutron EDM due to a vanishing Dirac trace when inserting it into the left diagram in Figure~\ref{fig:nEDM&Dmixing}, one can still impose a bound on the imaginary part of the scalar coefficient, $\mathrm{Im}[\hat{\epsilon}_S]$, by solving the RG equations together with the stringent limit on $\mathrm{Im}[\hat{\epsilon}_T]$ obtained from the neutron EDM. The RG running of these semi-leptonic coefficients could be found, \textit{e.g.}, in Ref.~\cite{Gonzalez-Alonso:2017iyc}. With the low-energy, the electroweak and the NP scale fixed respectively at $2~\mathrm{GeV}$, $m_Z$ and $1~\mathrm{TeV}$, and working at three-loop in QCD and one-loop in QED with both the bottom- and top-quark thresholds taken into account, the numerical result of the RG running below the electroweak scale reads~\cite{Gonzalez-Alonso:2017iyc}
\begin{align}\label{eq:RGEepsilon}
\left(
\begin{array}{c}
\hat{\epsilon}_S\\[0.1cm]
\hat{\epsilon}_T
\end{array}
\right)_{(\mu=2~\mathrm{GeV})}
= 
\left(
\begin{array}{cc}
 1.72 & -0.0242 \\[0.1cm]
 -2.17 \times 10^{-4} & 0.825
 \end{array}
\right)
\!\!\left(
\begin{array}{c}
\hat{\epsilon}_S\\[0.1cm]
\hat{\epsilon}_T
\end{array}
\right)_{(\mu=m_Z)}\,,
\end{align}
while above the electroweak scale it is given by~\cite{Gonzalez-Alonso:2017iyc}
\begin{align}\label{eq:RGEsmeft}
\left(
\begin{array}{c}
C^{(1)}_{\ell equ}\\[0.1cm]
C^{(3)}_{\ell equ}
\end{array}
\right)_{(\mu=m_Z)}
= 
\left(
\begin{array}{cc}
 1.20 & -0.185 \\[0.1cm]
 -0.00381 & 0.959
\end{array}
\right) 
\left(
\begin{array}{c}
C^{(1)}_{\ell equ}\\[0.1cm]
C^{(3)}_{\ell equ}
\end{array}
\right)_{(\mu=1~\rm{TeV})}\,.
\end{align}
The tree-level matching relations between the low-energy ($\hat{\epsilon}_{S,T}$) and the SMEFT ($C^{(1,3)}_{\ell equ}$) coefficients at the electroweak are given, respectively, by~\cite{Cirigliano:2009wk,Gonzalez-Alonso:2017iyc}
\begin{align}\label{eq:matchS}
\hat{\epsilon}_S =-\frac{[C^{(1)}_{\ell equ}]^*_{3321}}{2\sqrt{2}G_F V_{us}}\,, \qquad
\hat{\epsilon}_T =-\frac{[C^{(3)}_{\ell equ}]^*_{3321}}{2\sqrt{2}G_F V_{us}}\,,
\end{align}
where we have dropped the contribution from $C_{\ell edq}$ to $\hat{\epsilon}_S$, because, on the one hand, $C_{\ell edq}$ has no mixing with $C^{(3)}_{\ell equ}$ and, on the other hand, we are only interested in $C^{(1)}_{\ell equ}$ here. Combining Eqs.~\eqref{eq:RGEepsilon}--\eqref{eq:matchS} with the stringent bound $|\mathrm{Im}[\hat{\epsilon}_T](\mu_\tau)|\lesssim 4\times10^{-6}$ obtained from the neutron EDM in the ``single coefficient dominance'' assumption, we obtain
\begin{align}\label{eq:esfromnEDM}
|\mathrm{Im}[\hat{\epsilon}_S](\mu_\tau)|<2.3\times10^{-3}\,,
\end{align}
which is found to be comparable with that obtained from the $\tau\to K_S\pi\nu_\tau$ decays presented in subsection~\ref{sec:modelindepend}. This is due to the much smaller mixing effect of the scalar into the tensor operator, as can be clearly seen from Eqs.~\eqref{eq:RGEepsilon} and \eqref{eq:RGEsmeft}~\cite{Gonzalez-Alonso:2017iyc}. 

As the scalar interaction can also contribute to the $D^0-\bar{D}^0$ mixing via a double insertion of the operator $(\bar{\tau}_L\tau_R)(\bar{c}_Lu_R)$ into the right diagram shown in Figure~\ref{fig:nEDM&Dmixing}, we could also obtain another constraint on $\mathrm{Im}[\hat{\epsilon}_S]$ from this process. Following the same procedure as for $\mathrm{Im}[\hat{\epsilon}_T]$, and again in the ``single coefficient dominance'' assumption, we get the resulting short-distance Wilson coefficient
\begin{align}
C^\prime_2=-G_F^2\,\frac{m_\tau^2}{\pi^2}\,(V_{us}V_{cs}\,\hat{\epsilon}_S)^2\log\frac{\Lambda}{\mu_\tau}\,.
\end{align}
Choosing the scales $\Lambda=1~\mathrm{TeV}$ and $\mu_\tau=2~\mathrm{GeV}$, and using the latest global fit results for the $D^0-\bar{D}^0$ mixing parameters given by Eq.~\eqref{eq:x12}~\cite{Amhis:2019ckw}, we obtain the allowed $2\sigma$ range, $\mathrm{Im}[\hat{\epsilon}_S(\mu_\tau)]\in[-3.1, 1.6]\times10^{-4}$, which is found to be one order of magnitude stronger than that obtained from the neutron EDM (cf. Eq.~\eqref{eq:esfromnEDM}). For a simple estimate, let us choose $\mathrm{Im}[\hat{\epsilon}_S]=-3\times10^{-4}$ and plot the resulting $A_{CP}(s)$ as the red band in the right plot of Figure~\ref{fig:ACP}. It can be seen that the non-standard scalar contribution in this case is slightly smaller than the SM prediction for the $CP$ asymmetry in the angular distributions of $\tau\to K_S\pi\nu_\tau$ decays. However, when there exist extraordinary cancellations in the combinations $V_{ud}\,\mathrm{Im}[C^{(1)}_{\ell equ}]_{3311}+V_{us}\,\mathrm{Im}[C^{(1)}_{\ell equ}]_{3321}$ (for the neutron EDM) as well as $V_{cd}\,[C^{(1)}_{\ell equ}]_{3311}+V_{cs}\,[C^{(1)}_{\ell equ}]_{3321}$ (for the $D^0-\bar{D}^0$ mixing), the bound on $|\mathrm{Im}[\hat{\epsilon}_S]|$ could be significantly diluted, being comparable to the constraint from $\tau\to K_S\pi\nu_\tau$ decays presented in subsection~\ref{sec:modelindepend}, as can be seen from the right plot of Figure~\ref{fig:com}. We can therefore conclude that, once the bounds from the neutron EDM and the $D^0-\bar{D}^0$ mixing are taken into account, neither the scalar nor the tensor interaction can produce any significant effects on the $CP$ asymmetries (relative to the SM prediction) in the processes considered, especially under the ``single coefficient dominance'' assumption; nevertheless, when there exist extraordinary cancellations between the NP contributions, the non-standard scalar and tensor interactions can still produce observable effects on the $CP$ asymmetry in the angular distribution of $\tau\to K_S\pi\nu_\tau$ decays.

\section{Conclusion}
\label{sec:conclusion}

In this work, we have performed a detailed study of the $CP$ asymmetry in the angular distributions of $\tau\to K_S\pi\nu_\tau$ decays within a general EFT framework including four-fermion operators up to dimension-six. Such a bin-dependent $CP$-violating observable is more suitable than the decay-rate asymmetry, because the former, as usually measured in different $K\pi$ invariant-mass bins, can be made exempt from the uncertain inelastic phase of the $K\pi$ tensor form factor encountered in the latter, by setting the $K\pi$ invariant-mass intervals within the elastic region, where the explicit information of the tensor form factor is clear due to the Watson's final-state interaction theorem~\cite{Watson:1954uc}. It has been found that, besides the commonly considered scalar-vector interference, the tensor-scalar interference can also produce a non-vanishing $CP$ asymmetry in the angular distributions, provided that the couplings associated with the non-standard scalar and tensor interactions are complex.

For the numerical analyses, we have firstly updated our previous SM predictions~\cite{Chen:2020uxi} of the $CP$ asymmetries in the same four $K\pi$ invariant-mass bins as chosen by Belle, taking now into account the detector efficiencies of the Belle experiment~\cite{Bischofberger:2011pw}. Here we have used the more reliable dispersive representations instead of the Breit-Wigner parametrizations of the $K\pi$ vector, scalar, and tensor form factors, since the former warrant the properties of unitarity and analyticity, and contain a full knowledge of QCD in both the perturbative and non-perturbative regimes. Under the combined constraints from the $CP$ asymmetries measured in four $K\pi$ mass bins by Belle~\cite{Bischofberger:2011pw} as well as the branching ratio of $\tau^-\to K_S\pi^-\nu_\tau$ decay~\cite{Epifanov:2007rf}, the resulting bounds on the imaginary parts of the non-standard scalar and tensor couplings are given, respectively, by $\mathrm{Im}[\hat{\epsilon}_S]=-0.008\pm0.027$ and $\mathrm{Im}[\hat{\epsilon}_T]=0.03\pm0.12$. While our bound on $\mathrm{Im}[\hat{\epsilon}_S]$ is consistent with that obtained from Ref.~\cite{Bischofberger:2011pw}, the upper limit on $\mathrm{Im}[\hat{\epsilon}_T]$ is not competitive with that obtained from the neutron EDM and $D^0-\bar{D}^0$ mixing~\cite{Cirigliano:2017tqn}, which is attributed to the large uncertainties of the current Belle measurements of the $CP$ asymmetries~\cite{Bischofberger:2011pw}. Using the obtained best-fit values, we have also presented the distributions of the $CP$ asymmetries, finding that significant deviations from the SM prediction are possible in almost the whole $K\pi$ invariant-mass region. Therefore, the $CP$-violating angular observable considered here is an ideal probe of the non-standard scalar and tensor interactions. While being still plagued by large experimental uncertainties, the current constraints obtained in this work will be improved with more precise measurements from the Belle II experiment~\cite{Kou:2018nap}, as well as the future Tera-Z~\cite{Pich:2020qna} and STCF~\cite{Sang:2020ksa} facilities.

If the non-standard scalar and tensor operators contributing to the $\tau\to K_S\pi\nu_\tau$ decays are assumed to originate from a weakly-coupled heavy NP well above the electroweak scale, the $SU(2)_L$ invariance of the resulting SMEFT Lagrangian~\cite{Buchmuller:1985jz,Grzadkowski:2010es,Brivio:2017vri} would indicate that very strong limits on the imaginary parts of the non-standard scalar and tensor coefficients could also be obtained from the neutron EDM and the $D^0-\bar{D}^0$ mixing~\cite{Cirigliano:2017tqn}. To this end, we have exploited the maximum effects on the $CP$ asymmetries in the $\tau\to K_S\pi\nu_\tau$ decays, taking into account the constraints from the neutron EDM and the $D^0-\bar{D}^0$ mixing. It is found that, unless there exist extraordinary cancellations between the NP contributions, neither the scalar nor the tensor interaction can produce any significant effects on the $CP$ asymmetries (relative to the SM prediction) in the processes considered, especially under the ``single coefficient dominance'' assumption.

\section*{Acknowledgements}
This work is supported by the National Natural Science Foundation of China (NSFC) under Grant Nos.~12075097, 11675061, 11775092 and 11875327, the Fundamental Research Funds for the Central Universities under Grant Nos.~CCNU20TS007 and SYSU20191638, the
Natural Science Foundation of Guangdong Province, as well as the Sun Yat-Sen University Science Foundation.

\appendix
\renewcommand{\theequation}{A.\arabic{equation}}
\section*{Appendix: Dispersive representations of the \boldmath{$K\pi$} form factors}
\label{app:ff}
For the $K\pi$ vector and scalar form factors, one can find a brief summary of them in the appendix of Ref.~\cite{Chen:2020uxi}; for more details, the readers are referred to Refs.~\cite{Boito:2008fq,Boito:2010me} (for the vector form factor) and \cite{Jamin:2000wn,Jamin:2001zq,Jamin:2006tj} (for the scalar form factor), as well as the references therein. 

Let us now focus on the $K\pi$ tensor form factor. Unlike the previous two form factors, which are built with the relevant parameters determined phenomenologically, the $K\pi$ tensor form factor can be merely constructed from theory due to the lack of experimental data on the non-standard tensor interactions. Here we will make use of the once-subtracted dispersion relation~\cite{Cirigliano:2017tqn,Rendon:2019awg,Chen:2019vbr}
\begin{align}\label{eq:tff}
F_T(s) =F_T(0)\,\text{exp}\left\lbrace\frac{s}{\pi}\int_{s_{K\pi}}^\infty ds'\frac{\delta_T(s')}{s'(s'-s-i\epsilon)}\right\rbrace\,,
\end{align}
where the subtraction is fixed by the form factor at zero momentum transfer, $F_T(0)=\frac{\Lambda_2}{F_\pi^2}$, with $F_\pi$ being the physical pion decay constant and $\Lambda_2$ the low-energy constant of the effective Lagrangian of chiral perturbation theory ($\chi$PT) with tensor sources~\cite{Cata:2007ns,Mateu:2007tr}. Note that the coupling $\Lambda_2$ cannot be \textit{a priori} determined by $\chi$PT itself, but rather be inferred either from other low-energy constants using the short-distance constraint~\cite{Chen:2019vbr,Brodsky:1973kr,Lepage:1979zb,Lepage:1980fj} (see also Refs.~\cite{Husek:2020fru,Gonzalez-Solis:2019lze}) or from the lattice determination of the normalization $F_T(0)=0.417(15)$~\cite{Baum:2011rm}.\footnote{It should be noted here that our definition of the tensor form factor given by Eq.~\eqref{eq:THad} is different from that given in Ref.~\cite{Baum:2011rm}, where a factor $(M_{K}+M_{\pi})^{-1}$ has been inserted in order to make the tensor form factor dimensionless.} Here we will resort to the latter to obtain $\Lambda_2=(11.1\pm 0.4)~\mathrm{MeV}$.

\begin{figure}[t]
	\centering
	\includegraphics[width=0.62\textwidth]{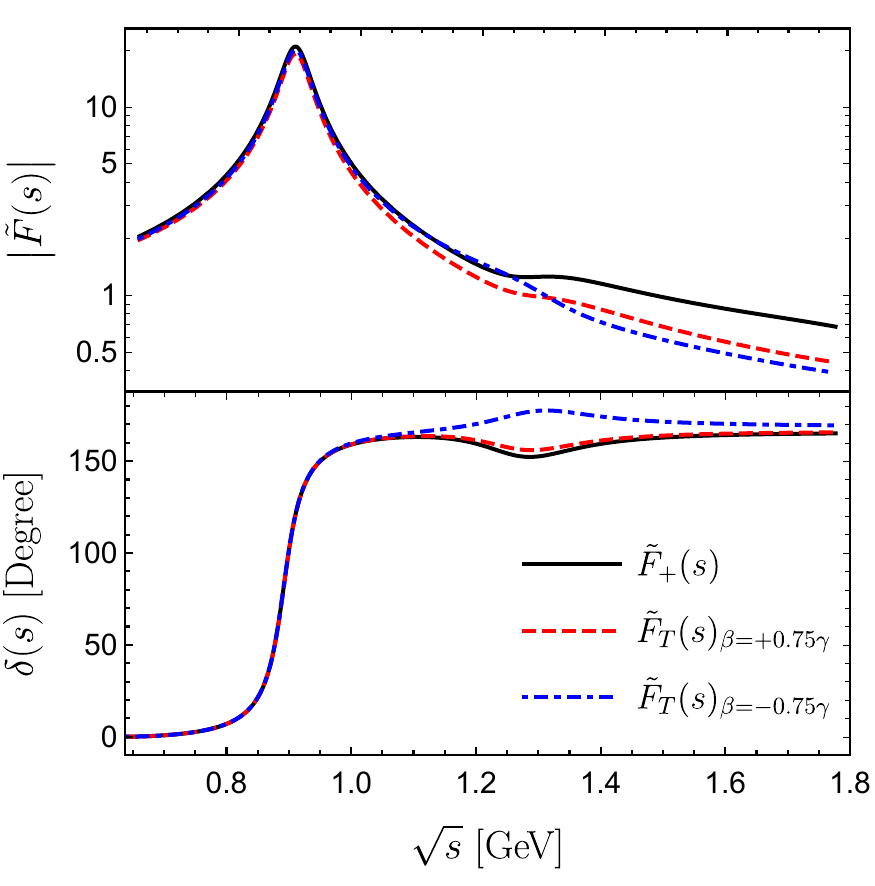}
	\caption{\small Moduli and phases of the normalized $K\pi$ vector ($\tilde{F}_{+}(s)$) and tensor ($\tilde{F}_{T}(s)$) form factors. The explicit expression of the vector form factor (black solid curves) is taken from Refs.~\cite{Boito:2008fq,Boito:2010me}, while that of the tensor form factor from Ref.~\cite{Chen:2019vbr}, with the red dashed and blue dot-dashed curves corresponding to the cases $\beta=+0.75\gamma$ and $\beta=-0.75\gamma$, respectively. \label{fig:FFs} }
\end{figure}

The tensor form-factor phase $\delta_T(s)$ in Eq.~\eqref{eq:tff} is calculated from the relation
\begin{align}\label{eq:tphase}
\tan\delta_T(s)=\frac{\mathrm{Im}[\tilde f_T(s)]}{\mathrm{Re} [\tilde f_T(s)]}\,,
\end{align}
where the explicit formula of the reduced tensor form factor $\tilde f_T(s)$ has been given by Eq.~(4.18) of Ref.~\cite{Chen:2019vbr}, which is derived in the context of R$\chi$T with the vector resonances $K^\ast(892)$ and $K^\ast(1410)$ included as explicit degrees of freedom~\cite{Chen:2019vbr}. It should be noted that, in the context of a vector-meson-dominance picture, the $K\pi$ vector and tensor form factors are both dominated by the same vector resonances~\cite{Cirigliano:2017tqn}. In order to show the relationship between the $K\pi$ vector and tensor form factors, we plot in Figure~\ref{fig:FFs} both the moduli and the phases of the two normalized form factors $\tilde{F}_{i}=F_i(s)/F_i(0)$, with $i=+,T$. It can be seen that $\delta_T(s)=\delta_+(s)$ in the elastic region, validating therefore the Watson's final-state interaction theorem~\cite{Watson:1954uc}. In the inelastic region, on the other hand, $\delta_T(s)$ and $\delta_+(s)$ start to behave differently due to the different relative weights of the two resonances in the tensor and vector form factors, which are characterized by the two mixing parameters $\beta$~\cite{Chen:2019vbr} and $\gamma$~\cite{Boito:2008fq,Boito:2010me}, respectively. Although the parameter $\beta$ cannot be determined directly from data for the moment, the ratio $\beta/\gamma$ can be estimated from the large-$N_c$ patterns of the correlators derived in Ref.~\cite{Cata:2008zc}, which eventually leads to the relation $\beta=\pm0.75\gamma$~\cite{Chen:2019vbr}. 

In this work, as a conservative estimate, we will take the range varied between the positive (the red dashed curves with $\beta=+0.75\gamma$) and negative (the blue dot-dashed curves with $\beta=-0.75\gamma$) inputs of $\beta$ as the systematic uncertainty of the tensor form factor. As Eq.~\eqref{eq:tphase} is valid only in the region from the threshold $s_{K\pi}$ to $m_\tau^2$, further information of the tensor form factor in the higher-energy region is required to compute the dispersive integral. This is, however, unrealistic due to the lack of experimental data on the non-standard tensor interactions. In fact, one can only estimate the phase in the higher-energy region based on the asymptotic behaviour of the tensor form factor at the large $K\pi$ invariant-mass squared $s$~\cite{Brodsky:1973kr,Lepage:1979zb,Lepage:1980fj}. As detailed in Ref.~\cite{Chen:2019vbr}, by introducing different choices of the cut-off $s_{cut}$ as well as different asymptotic values $n_T\pi$ of the phase in the inelastic region, it is found that the modulus of the normalized tensor form factor is almost insensitive to the choice of $s_{cut}$ when fixing $n_T=1$, while it becomes rather sensitive to the choice of $n_T$ when fixing $s_{cut}=4~\mathrm{GeV}^2$, especially in the higher-energy region. This implies that the once-subtracted dispersive representation given by Eq.~\eqref{eq:tff} is not optimal, as is generally expected. Nevertheless, the lack of experimental data sensitive to the tensor form factor makes it impossible to increase the number of subtractions for the moment~\cite{Chen:2019vbr}. For recent discussions about the two-hadron tensor form factors, the readers are also referred to Refs.~\cite{Hoferichter:2018zwu,Husek:2020fru,Shi:2020rkz,vonDetten:2021euh}. 

\bibliographystyle{JHEP}
\bibliography{reference}

\end{document}